\documentclass{article}
\PassOptionsToPackage{svgnames}{xcolor}
\usepackage{lmodern}
\usepackage{geometry}
\usepackage{anyfontsize}
\usepackage{microtype} 
\usepackage{amsmath,amssymb,amsopn,amsthm,mathrsfs,mathtools}
\usepackage{graphicx}
\usepackage{bm}
\usepackage{citesort}
\usepackage{appendix}
\usepackage{enumitem}
\usepackage[affil-it]{authblk}
\usepackage{tikzit}
\usepackage{tcolorbox}
\usepackage{cancel}
\usepackage{comment}
\usepackage{cite}
\usepackage{ulem}
\usepackage{tabu}
\usepackage{multirow}
\usepackage{caption}
\usepackage{subcaption}
\usepackage{setspace}

\newcommand{\abrac}[1]{\left\langle #1 \right\rangle}

\newcommand{\tepsilon}{\bm{\tilde{\varepsilon}}}

\newcommand{\tepsilonu}[1]{\bm{\tilde{\varepsilon}}_{#1}(\bm{u})}

\newcommand{\divu}{\text{div}\bm{u}}

\newcommand{\uD}{\bm{u}\!\cdot\!\nabla}

\newcommand{\Lmu}{\mathcal{L}_{\mu} }
\newcommand{\iLmu}{\mathcal{L}_{\mu}^{-1}}

\newcommand{\LM}{\mathcal{L}_{\mathcal{M}}}
\newcommand{\iLM}{\mathcal{L}_{\mathcal{M}}^{-1}}

\newcommand{\Id}{\text{Id}}
\newcommand{\St}{\text{St}}

\newcommand{\KrPhi}{\bm{\Phi}}

\newcommand{\vdx}[1]{v_{#1}\partial_{x_{#1}}}
\newcommand{\dx}[1]{\partial_{x_{#1}}}

\newcommand{\parent}[1]{\left( #1\right)}

\newcommand{\sqbrac}[1]{\left[#1 \right]}

\newcommand{\trho}{\tilde{\rho}}

\newcommand{\tu}{\tilde{u}}

\newcommand{\ttheta}{\tilde{\theta}}

\usetikzlibrary{shapes,arrows}
\tikzstyle{line}=[draw]
\tikzstyle{startstop} = [rectangle, rounded corners, minimum width=3cm,
    minimum height=1cm,text centered, draw=black, fill=red!30]

\newcommand{\Xtens}[7]{\bm{\mathrm{X}}^{#1\, #2\, #3\, #4}_{\phantom{#1}#5 #6 #7}}
\newcommand{\Ltens}[6]{\bm{\mathrm{L}}^{#1\, #2\, #3\, #4}_{\phantom{#1}#5 \ #6}}
\newcommand{\Ytens}[8]{\bm{\mathrm{Y}}^{#1\, #2\, #3\, #4}_{ #5 #6 #7 #8}}
\DeclareMathOperator*{\argmin}{arg\,min}

\newtheorem{theorem}{Theorem}[section]
\newtheorem{proposition}{Proposition}[section]
\newtheorem{corollary}[theorem]{Corollary}
\newtheorem{lemma}[theorem]{Lemma}
\newtheorem{remark}{Remark}

\tcbset{boxrule=0.5pt,
        arc=2mm,
        top=1mm,bottom=1mm,right=1mm,left=1mm}

\title{Extensions to the Navier-Stokes-Fourier Equations for Rarefied Transport: Entropy Variable Formulation, Heat Transfer and Dispersion of a Non-Linear Entropy Stable Closure}
\author{Frimpong A. Baidoo, Narayana R. Aluru, Thomas J.R. Hughes, Michael R.A. Abdelmalik}
\date{November 2026}

\begin{document}

\maketitle
\begin{abstract}
We assess the non-linear version of the Entropy Stable Extension to the Navier-Stokes-Fourier equations which were originally derived in \cite{Baidoo2026}. Using an identity for the derivative of the inverse of linearized collision operator, we show that the Entropy Stable Extension takes the form of a fourth-order quasi-cubic system of equations when written in entropy variables. Furthermore, we find that the quasi-linear contributions to the extension retain the highest order derivative contributions and the entropy stability of the full extension. 
This quasi-linear approximation is applied to the stationary heat transfer between parallel plates where we find that a specification of the boundary conditions that interpolates between the continuum and collisions limits can yield temperature profiles that better agree with DSMC predictions than Navier-Stokes-Fourier equations with Smoluchowski boundary conditions, while also giving a heat flux prediction that better agrees with the Sherman-Lees heat flux than the Navier-Stokes-Fourier prediction. We conclude with a brief study of dispersion of plane waves according to the linearized equations in one dimension where we show that all frequency modes are stable and relate this to entropy stability.  
\end{abstract}

\section{Introduction}\label{sec:Introduction}

In flow regimes where the mean free path of the gas is comparable to a macroscopic reference length scale of the domain under consideration, equations such as the Euler and the Navier-Stokes-Fourier (NSF) equations fail to accurately describe the flow characteristics of a gas. Such flow regimes are common at high altitudes, in near vacuum conditions and in micro and nanoscale gas flows. 
The Boltzmann transport equation overcomes this failure to model these types of rarefied flows by describing the gas in terms of a distribution function that evolves through phase space as a result of interparticle collisions. In principle, this means that it can describe gaseous behavior in all flow regimes. However, directly computing the Boltzmann equation presents a formidable challenge due to the need to discretize an unbounded phase space velocity domain and a non-linear, non-local collision operator. An alternative to working with the Boltzmann equation directly is to solve a finite number of equations derived by taking moments of the Boltzmann equation and closing the resulting system. The derivation of such moment equations from the Boltzmann equation has a rich history. This includes the Chapman-Enskog expansion \cite{chapman1990mathematical, Enskog1917} which closes a system of equations for mass, momentum and energy (5 moments) of the gas and from which is derived the Euler, NSF and the Burnett equations\cite{Burnett1935}. 
It also includes Grad's moment equations \cite{grad1949}, which closes a system in which the mass, momentum and energy equations are a strict subset, as well as the regularized variants that improve upon it such as the Regularized 13 moment (R13) equations \cite{Struchtrup2003} and the Regularized 26 moment (R26) equations \cite{GU2009}.   

In \cite{Baidoo2026}, a novel extension to the NSF equations was derived from the Boltzmann equation through the Variational Multiscale Moment Method framework. This Entropy Stable Extension to the Navier-Stokes-Fourier (ESE) is a fourth-order system of conservation equations meant to overcome the well-known shortcomings of the Burnett equations, such as the violation of the second law of thermodynamics \cite{Comeaux1995}, while serving the same role of providing a 5 moment macroscopic model of rarefied gas flow that improves upon the NSF equations. In \cite{Baidoo2026} and later on in \cite{Baidoo2026_HD}, the linearized version of the ESE equations were shown to reproduce many important flow characteristics across a wide range of Knudsen regimes for the Couette flow, Stationary Heat Transfer and Poiseuille flow benchmark problems. Not much attention was paid to the non-linear version of the ESE equations beyond showing that they were entropy stable. In this manuscript, we seek to take a deeper look at the non-linear ESE equations and to further test its effectiveness for rarefied gas dynamics.

The manuscript is structured as follows. In Section~\ref{sec:BoltzmannEquation}, we introduce the Boltzmann equation and briefly describe some important properties of the collision operator. We then introduce the linearized collision operator in Section~\ref{sec:LinearizedCollisionOperator} and discuss the properties needed to derive closed systems of conservation equations. This includes proving in Subsection~\ref{subsec:derivLinCol}, a characterization of the derivative of the inverse of the linearized collision operator that is, to our knowledge, novel. This characterization is eventually applied when describing the ESE equations using entropy variables. In Section~\ref{sec:VMS}, we briefly describe the Variational Multiscale Moment Method and how the Euler, Navier-Stokes-Fourier (NSF) and ESE arise from this framework. We then describe these conservation equations in terms of entropy variables in Section~\ref{sec:EntropyVariables} shedding some light on the relationship between their symmetric structure and entropy stability. Section~\ref{sec:quasilinearESE} describes the quasi-linear approximation to the ESE equations, an important subset of the ESE equations that contains the highest order derivative contributions and retains the entropy stability of the full ESE equations. The fluid parameters pertaining to Hard Spheres and Maxwell Molecules are also tabulated here. Section~\ref{sec:StationaryHeat} applies the quasi-linear approximation to the Stationary Heat transfer problem. Solving this problem requires specifying boundary conditions for the quasi-linear ESE equations. We use slip functions to specify a set of boundary conditions that look like the Smoluchowski jump boundary conditions in the continuum limit but mimic the collisionless solution for large Knudsen numbers. We compare the ESE solutions obtained with these boundary conditions to solutions of the NSF with Smoluchowski boundary conditions, Direct Simulation Monte Carlo and Sherman-Lees. Finally, Section~\ref{sec:Dispersion} studies the dispersion in the one-dimensional linearized ESE equations in comparison to those of the NSF equations. We then draw conclusions in Section~\ref{sec:Conclusions}. The Appendix briefly discusses the ESE equations in relation to the BGK collision operator, describes how the macroscopic closures are obtained, details the parameters used to generate DSMC solutions for the Stationary Heat Transfer problem and the procedure by which the slip functions used for the boundary conditions were obtained.

\section{The Boltzmann Equation}\label{sec:BoltzmannEquation}
The evolution of a dilute gas of monoatomic particles can be described by the Boltzmann transport equation 
\begin{align}\label{eq:BoltzmannEqn}
    \text{St}\,\partial_t F + v_i \partial_{x_i} F =  \frac{1}{\epsilon}\parent{\epsilon J(F, F)} =:  \frac{1}{\epsilon}Q(F, F)
\end{align}
where $F(t, \boldsymbol{x}, \boldsymbol{v})$ is the distribution function of gas in phase space $\mathbb{R}_{\bm{x}}^3\times\mathbb{R}_{\bm{v}}^3$, the \textit{Knudsen number} $\epsilon$ is the ratio of the mean free path of the gas and a reference macroscopic length scale $L$, the \textit{kinetic Strouhal number} $\text{St}$ is the ratio of the macroscopic length scale to the product of thermal speed of sound of the gas and a reference macroscopic timescale. In the above equation and throughout the text, we have employed the Cartesian tensor convention where repeated indices imply summation up to the number of spatial dimensions (i.e. three).

The \textit{collision operator} $Q(F, F)$ on the right hand side of \eqref{eq:BoltzmannEqn} accounts for the effect of interparticle collisions on evolution of the distribution function through phase space. It is a non-local function of $\bm{v}$ and a local function of $t$ and $\bm{x}$. In this manuscript, we shall pay particular attention to the bilinear collision operators, which take form \small
\begin{align}\label{eq:BilinearOperator}
    Q(G, H)(\bm{v}) &= \frac{1}{2}\int_{\mathbb{R}^3}\! \int_{\mathbb{S}^2}\!\big(G(\bm{v}')H(\bm{v}_*') +G(\bm{v}'_*)H(\bm{v}') - G(\bm{v})H(\bm{v}_*)-G(\bm{v}_*)H(\bm{v})\big)\mathfrak{B}\!\left(|\bm{v_* - \bm{v}}|,\left|\hat{\bm{\omega}}\cdot\!\frac{(\bm{v}_*-\bm{v})}{|\bm{v_* - \bm{v}}|}\right|\right)\, d\hat{\bm{\omega}} d\bm{v}_*
\end{align}\normalsize
where $\bm{v}' = \bm{v} + \big(\hat{\bm{\omega}}\cdot(\bm{v}_*-\bm{v})\big)\hat{\bm{\omega}} $, $\bm{v}_*' = \bm{v}_* - \big(\hat{\bm{\omega}}\cdot(\bm{v}_*-\bm{v})\big)\hat{\bm{\omega}}$ and $\mathfrak{B}$ is the collision kernel which encodes interparticle potential.  The domain of the collision operator $\mathscr{D}(Q)$ is closed under translations and rotations of the phase-space velocity $\bm{v}$ and a subset of $\mathscr{L}^1_2(\mathbb{R}^3_{\bm{v}})\cap\mathscr{L}\ln\mathscr{L}(\mathbb{R}^3_{\bm{v}})$ where $\mathscr{L}_2^1(\mathbb{R}_{\bm v}^3) := \{ G : \int_{\mathbb{R}^3} (1 + |\bm v|^2) G(t, \bm x, \bm v)\, d\bm v < \infty \}$ and $\mathscr{L}\ln \mathscr{L}(\mathbb{R}_{\bm v}^3):= \{ G : \int_{\mathbb{R}^3}G \ln(G) \, d\bm v < \infty\} $. We would like to note that our definition of the collision operator $Q$, differs from the usual definition of the collision operator $J = Q / \epsilon$ (for example, see Equation 3 of \cite{Aoki2017}). This, together with the rescaling $t = \St\, \tilde{t}$ in time, is done in order to pull out the Knudsen and Strouhal numbers without non-dimensionalizing the equation. Non-dimensionalization of equations will only be performed when working with a particular problem through the use of reference values relevant to the problem in question. 

Two common collision kernels for the bilinear operator \eqref{eq:BilinearOperator} are:
\begin{enumerate}
    \item \textit{Hard spheres}: 
    \begin{equation}\label{eq:HardSpheresCrossSection}
        \mathfrak{B}\!\left(|\bm{v_* - \bm{v}}|,\left|\hat{\bm{\omega}}\cdot\!\frac{(\bm{v}_*-\bm{v})}{|\bm{v_* - \bm{v}}|}\right|\right) =\frac{\epsilon d_m^2}{2 m}|\hat{\bm{\omega}}\cdot\!(\bm{v}_*-\bm{v})| = \frac{1}{2\pi\rho_0 L\sqrt{2} }\,|\hat{\bm{\omega}}\cdot\!(\bm{v}_*-\bm{v})|
    \end{equation}
     $d_m$ is the diameter of a constituent particle, $m$ is the particle mass, $\rho_0$ is a reference density. The last equality was obtained with the following formula for the mean free path $\ell = \epsilon L = \frac{m}{d_m^2 \rho_0 \pi \sqrt{2}}$.
    \item \textit{Maxwell molecules}
    :
    For a gas with an interparticle potential given by $V(r) = V_0\left(\frac{d_m}{r} \right)^4$, the collision kernel takes the form 
    {
\begin{equation}\label{eq:MaxMolCrossSection} 
\begin{aligned}
\mathfrak{B}\!\left(|\bm{v_* - \bm{v}}|,\left|\hat{\bm{\omega}}\cdot\!\frac{(\bm{v}_*-\bm{v})}{|\bm{v_* - \bm{v}}|}\right|\right)  &= \frac{\epsilon d_m^2}{2m}\sqrt{\frac{V_0}{m}}\,\mathfrak{F}\left(\vartheta\right) = \frac{1}{2\pi \rho_0 L }\sqrt{\frac{V_0}{2m}}\,\mathfrak{F}\left(\vartheta\right)    
\end{aligned}
    \end{equation} 
    }
where the scattering angle $\vartheta$ is defined by $\cos\vartheta = 1-2\left|\hat{\bm{\omega}}\cdot\!\frac{(\bm{v}_*-\bm{v})}{|\bm{v_* - \bm{v}}|}\right|^2$ and $$\mathfrak{F}(\vartheta) = \frac{2}{\sin(\vartheta)\sin^2(2\psi)}\left(\frac{d\vartheta}{d\psi} \right)^{-1}$$ The angle $\psi$ defined implicitly by the relation $\vartheta(\psi) = \pi - 2\sqrt{\cos(2\psi)} \,K\big(\sin(\psi)\big)$ where $K$ is the complete elliptic integral of the first kind.

\end{enumerate}
\begin{remark}
    The exposition in the text is greatly simplified by focusing on bilinear operators like the above. However, there exist other commonly used classes of collision operators such as the  Bhatnagar-Gross-Krook (BGK) model \cite{bhatnagar:1954hc} and the Ellipsoidal Statistical model \cite{Holway1966}. In the appendix, we will very briefly describe results as they relate to the BGK model.
\end{remark}
Velocity moments of the distribution function correspond to macroscopic properties of the gas. The moments of particular interest to us are: 
\begin{subequations}\label{eq:macroparam}
\begin{align}
\rho &=\int_{\mathbb R^D} F(t, \bm x, \bm v) \, d\bm v  = \left\langle 1,\, F \right\rangle \label{eq:mass}\\ 
 u_i&= \frac{1}{\rho}\int_{\mathbb R^D} v_i\, F(t, \bm x, \bm v) \, d\bm v =  \frac{1}{\rho}\left\langle v_i , \, F \right\rangle  \label{eq:bulkVelocity}\\
 \theta &=\frac{1}{3\rho}\left\langle |\bm v - \bm u|^2,\, F \right\rangle  \label{eqq:temperature}\\
\sigma_{ij}&= -\left\langle\Big( (v_i - u_j)(v_j - u_j) - \frac{1}{3}|\bm v - \bm u|^2\delta_{ij}\Big),\, F\right\rangle 
\label{eq:deviatoricStress}\\
q_i &=\frac{1}{2}\left\langle  \left(|\bm v - \bm{u}|^2 (v_i - u_i) - 5\theta\, (v_i-u_i)\right), \,F \right\rangle 
\label{eq:heatFlux}
\end{align}
\end{subequations}
where $\rho(t,\bm{x})$ is the mass density, $\bm{u}(t,\bm{x}) = (u_1, u_2, u_3)^\mathsf{T}$ is the bulk velocity, $\theta(t,\bm{x}) = k_B T/m$ is the temperature rescaled by the Boltzmann constant $k_B$, the deviatoric stress is $\sigma_{ij}(t,\bm{x})$ and the heat flux is $q_i(t,\bm{x})$.

\subsection{Properties of the Collision Operator}
All collision operators have to satisfy certain properties in order to be physically realistic. A more detailed treatment of these properties can be found in \cite{cercignani1988}. The properties of particular interest to us are 
\begin{enumerate}
\item \textit{Translational and Rotational Invariance}: The collision operator commutes with the translation and rotation operators $\mathcal{T}_{\bm{U}}[g] := g(\bm{v} - \bm{U})$ and  $\mathcal{T}_{\mathbf{O}} := g(\mathbf{O}^{\mathsf{T}}\bm{v})$ for any vector $\bm{U}\in\mathbb{R}^3$ and orthogonal matrix $\mathbf{O}\in \mathbb{R}^{3\times3}$, i.e.
\begin{align}\label{eq:TransRotInvariance}
 \mathcal{T}_{\bm{U}}[Q(G, G)] = Q\big(\mathcal{T}_{\bm{U}}[G], \mathcal{T}_{\bm{U}}[G]\big) && \mathcal{T}_{\mathbf{O}}[Q(G, G)] = Q\big(\mathcal{T}_{\mathbf{O}}[G], \mathcal{T}_{\mathbf{O}}[G]\big)   
\end{align}
    \item \textit{Collision invariance}: With the space of collision invariants defined as $\mathscr{I}:= \text{span}\{1, v_1, v_2, v_3, |\bm{v}|^2 \}$ one can show that collision operators of the form \eqref{eq:BilinearOperator} are such that 
    $$\left\langle \bar{m}(\bm{v}),\, Q(G, G) \right\rangle=0 \qquad\forall G \in \mathscr{D}(Q)\quad  \Longleftrightarrow \quad\bar{m} \in \mathscr{I}$$
   A consequence of collision invariance is that testing the Boltzmann equation \eqref{eq:BoltzmannEqn} with $\bar{m} = 1, v_i$ and $\frac{|\bm{v}|^2}{2}$ of the Boltzmann equation \eqref{eq:BoltzmannEqn} yields a system of conservation equations in mass, momentum and energy of the gas
\begin{subequations}
\label{eq:macroconslaws}
\begin{align}
\text{St }\partial_t\rho + \partial_{x_i} [\rho u_i]&=0; \label{eq:masslaw}\\
\text{St }\partial_t[\rho u_j] + \partial_{x_i} [\rho {u}_i{u}_j+ \rho \theta\delta_{ij} - \sigma_{ij}]&=0;\label{eq:momentumlaw}\\
\text{St }\partial_t\left[\frac{1}{2}\rho|\bm u|^2+\frac{3}{2}\rho \theta\right] + \partial_{x_i} \left[\left(\frac{1}{2}\rho|\bm u|^2 u_i + \frac{3}{2} \rho\theta \right)u_i+ \rho\theta u_i- \sigma_{ij} u_j + q_i\right]&=0\label{eq:energylaw};
\end{align}
\end{subequations}
\normalsize The above system of five equations contains thirteen unknowns and would require closure relations in order to be used directly. It is therefore necessary to have a framework through which closure relations for $\sigma_{ij}$ and $q_{i}$ are derived in terms of $\rho,\, u_i$ and $\theta$ and their derivatives.

\item \textit{Entropy Dissipation and Equilibrium}: For all $G \in \mathscr{D}(Q)$, we have that
    $$ \left\langle \ln(G),\, Q(G,G)\right\rangle \le 0$$
    with equality attained if and only if $Q(G,G) = 0$. This property leads to Boltzmann's H-theorem when we test \eqref{eq:BoltzmannEqn} by $\ln(F)$ and integrate in velocity to obtain the entropy dissipation inequality for the Boltzmann equation\footnote{Note that due to conservation of mass \eqref{eq:masslaw}, we could replace $\ln(F) - 1$ with $\ln(F) - c$ for an arbitrary constant $c$.}
\begin{align}\label{eq:entdissip}
    \text{St}\,\partial_t\! \left\langle \ln\left({F}\right)- 1, \,F \right\rangle
 +
 \partial_{x_i}\! \left\langle \ln\left({F}\right)-1, \,  v_i\,F\right\rangle 
 = 
 \frac{1}{\epsilon}\left\langle \ln\left({F}\right),\,  Q(F,F)\, \right\rangle\leq 0
\end{align}
In combination with the collision invariance property, the above inequality says that equilibrium in achieved if and only if $\ln(F) \in \mathscr{I}$. This defines a family of distribution functions referred to as \textit{Maxwellians} which can, without loss of generality, be parametrized as 
\begin{equation}\label{eq:MaxwellianDef}
    \mathcal{M}_{\varrho, \bm{U}, \vartheta}(t, \bm{x}, \bm{v}) := \frac{\varrho(t, \bm{x})}{(2\pi\,\vartheta(t, \bm{x}))^{\frac{3}{2}}}\exp\left(- \frac{|\bm{v} - \bm{U}(t, \bm{x}) |^2}{2\,\vartheta(t, \bm{x})} \right)
\end{equation}
Due to the requirement that distributions be non-negative and integrable over $\mathbb{R}^3_{\bm{v}}$, we observe that $\varrho$ must be non-negative and $\vartheta$ positive. In situations where the conservation moments $(\varrho, \bm{U}, \vartheta)$ of the Maxwellian are contextually unambiguous, we will take to writing $\mathcal{M}_{\varrho, \bm{U}, \vartheta}$ as simply $\mathcal{M}$. For a given distribution $F(t,\bm{x}, \bm{v})$, we can derive a corresponding \textit{self-consistent Maxwellian} $\mu(F)$ which is the unique Maxwellian that shares that same conservation moments $(\rho, \bm{u}, \theta)$ as $F$. In other words, 
\begin{equation}\label{eq:selfconsMaxDef}
\mu(F) := \frac{\rho}{(2\pi\,\theta)^{\frac{3}{2}}}\exp\left(- \frac{|\bm{v} - \bm{u}|^2}{2\,\theta} \right)
\end{equation} 
where
\begin{equation}\label{eq:selfconsMaxDef2}
\begin{pmatrix}
\rho
\\ 
\rho\, \bm u
\\
\rho |\bm u|^2 + 3\rho\theta
\end{pmatrix}
\ :=\
\begin{pmatrix} 
\left\langle1, \, F \right\rangle
\\ 
\left\langle \bm v,\, F \right\rangle
\\ 
\left\langle |\bm v|^2,\, F  \right\rangle
\end{pmatrix} 
\ =\ 
\begin{pmatrix} 
\left\langle 1,\, \mathcal \mu(F)  \right\rangle
\\ 
\left\langle \bm v,\, \mu(F) \, \right\rangle
\\ 
\left\langle |\bm v|^2,\, \mu(F)  \right\rangle
\end{pmatrix}
\end{equation}

\end{enumerate}
Observe that $\mu(F) = F$ if and only if $F$ is a Maxwellian.

\section{Linearized Collision Operator}\label{sec:LinearizedCollisionOperator}
Because Maxwellians serve as equilibrium functions to the collision operators, the behavior of collision operators when a distribution function is close to a Maxwellian $\mathcal{M}(t, \bm{x}, \bm{v})$ is of particular interest. This motivates the linearized collision operator with respect to $\mathcal{M}$ 
\begin{align}\label{eq:LinColOpDef}
\mathcal{L}_{\mathcal{M}}[g] &:= \frac{1}{\mathcal{M}} \,\left. \frac{d}{d\eta}Q\Big(\mathcal{M}(1 + \eta g), \mathcal{M}(1 + \eta g)\Big)  \right|_{\eta=0} = \frac{2}{\mathcal{M}}Q(\mathcal{M}, \mathcal{M}g)\\
&= \int_{\mathbb{R}^3}\! \int_{\mathbb{S}^2}\! \mathcal{M}(\bm{v}_*)\Big(g' +g'_* - g -g_*\Big)\mathfrak{B}\!\left(|\bm{v_* - \bm{v}}|,\left|\hat{\bm{\omega}}\cdot\!\frac{(\bm{v}_*-\bm{v})}{|\bm{v_* - \bm{v}}|}\right|\right)\, d\hat{\bm{\omega}} d\bm{v}_*\nonumber\\
    &=:\, \mathtt{L}[\mathcal{M},\, g] \nonumber
\end{align}
In the above, we have used the identity $\mathcal{M}(\bm{v}_*)\mathcal{M}(\bm{v}) = \mathcal{M}(\bm{v}_*') \mathcal{M}(\bm{v}')$ and employed the commonly used convention $g' := g(\bm{v}')$, $g'_* := g(\bm{v}'_*)$, $g := g(\bm{v})$ and $g_* := g(\bm{v}_*)$ for notational brevity. The bilinear operator is written in the form $\mathtt{L}[\mathcal{M}, \,g]$ in the last line so as to draw attention to the fact that the linearized collision operator acts linearly on both $\mathcal{M}$ and $g$. 
In order to apply the Variational Multiscale moment method described in the next section, the linearized collision operator has to satisfy the following properties:
\begin{enumerate}
    \item \textit{Rotational invariance}: Due to \eqref{eq:TransRotInvariance}, for a Maxwellian $\mathcal{M} = \mathcal{M}_{\varrho, \bm{U}, \vartheta}$ and an arbitrary orthogonal matrix $\bm{\mathrm{O}} \in \mathbb{R}^{3\times3}$ one can show that 
\begin{align}\label{eq:LinRotInvariance}
    \LM[g(\bm{v}_*)]\big(\bm{\mathrm{O}}^{\mathtt{T}}(\bm{v} - \bm{U})+\bm{U}\big) = \LM\big[g\big(\bm{\mathrm{O}}^{\mathtt{T}}(\bm{v}_* - \bm{U})+\bm{U}\big)\big](\bm{v})
\end{align}
    This particular property is important for analyzing tensors whose entries are integrals made up of $\mathcal{M}$, polynomials in $\bm{c} := (\bm{v}-\bm{U})$ and $\mathcal{L}_{\mathcal{M}}$ acting on polynomials of $\bm{c}$, for example 
    $${J}_{i_1\,i_3\,i_2}^{j_1\,j_2\,j_3} = \abrac{c_{i_1}\mathcal{L}_{\mathcal{M}}\left[c_{i_2} c_{i_3}\right],\, \mathcal{M}\, \mathcal{L}_{\mathcal{M}}\left[c_{j_1}\mathcal{L}_{\mathcal{M}}\left[|\bm{c}|^2 c_{j_2} c_{j_3}\right]\right] }$$
As a consequence of \eqref{eq:LinRotInvariance}, tensors such as ${J}_{i_1\,i_3\,i_2}^{j_1\,j_2\,j_3}$ will be isotropic tensors i.e. they are unchanged by a rotational change of coordinates. This greatly simplifies the form in which the derived closures take.  For example, any rank 2 isotropic tensor is necessarily a multiple of the Kronecker delta tensor $\delta_{ij}$. What's more, using $\mathbf{O} = -\text{Id}$, we can show that if the integrand for such a tensor has an odd number of indices, the entries of that tensor will be zero. 
    \item \textit{Self-Adjointness}: The linearized operator $\mathcal{L}_{\mathcal{M}}$ is self-adjoint with respect to the inner product of the weighted space $\mathscr{L}^2(\mathcal{M}d\bm{v}) := \{g : \abrac{g,\, \mathcal{M} g} \le \infty \}$. In other words,
    $$\abrac{g, \,\mathcal{M}\,\mathcal{L}_{\mathcal{M}}[h]} = \abrac{\mathcal{L}_{\mathcal{M}}[g],\, \mathcal{M}\, h}$$
 \item \textit{Collision Invariance}: 
    $$\mathcal{L}_{\mathcal{M}}[\bar{g}] = 0 \ \Longleftrightarrow \ \bar{g} \in \mathscr{I}$$
    For a  linearized collision operator of the form \eqref{eq:LinColOpDef}, the reverse direction of this equivalence is easily derived by observing that the conservation relation $\bar{g}'+ \bar{g}_*' = \bar{g} + \bar{g}_*\,$ holds for the collision invariants. The forward direction is a bit more subtle but proofs can be found in works such as \cite{Cercignani1990, alonsogamba2022}. 
    In combination with self-adjointness, collision invariance implies that the range of the linearized collision operator is orthogonal to the space of collision invariants because
    $$\abrac{\bar{g}, \,\mathcal{M}\,\mathcal{L}_{\mathcal{M}}[h]} = \abrac{\mathcal{L}_{\mathcal{M}}[\bar{g}],\, \mathcal{M}\, h} = 0$$
     This motivates decomposing $\mathscr{L}^2(\mathcal{M}d\bm{v})$ into the space of collision invariants $\mathscr{I}$ and the orthogonal complement to the space of collision invariants $\mathscr{I}^{\perp_{\mathcal{M}}}$, i.e. $\mathscr{L}^2(\mathcal{M}d\bm{v}) = \mathscr{I}\oplus\mathscr{I}^{\perp_{\mathcal{M}}} $ 
    \item \textit{Negative semi-definiteness}: 
    $$\abrac{\mathcal{L}_{\mathcal{M}}[g],\, \mathcal{M}\, g} \le 0 $$
    \item \textit{Fredholm Alternative}: The linearized collision operator $\mathcal{L}_{\mathcal{M}}$ restricted to $\mathscr{I}^{\perp_{\mathcal{M}}}$ is invertible. We can extend this inverse to the rest of $\mathscr{L}^2(\mathcal{M}d\bm{v})$ by making the space of collision invariants $\mathscr{I}$ its kernel. Abusing notation, we shall also refer to this extension of the inverse as $\mathcal{L}_{\mathcal{M}}^{-1}$. Notice that the inverse linearized operator $\mathcal{L}_{\mathcal{M}}^{-1}$ will also inherit rotational invariance, self-adjointness and negative semi-definiteness from the linearized operator $\mathcal{L}_{\mathcal{M}}$.
    If we denote the orthogonal projection onto $\mathscr{I}$ by $\Pi_{\mathcal{M}}$, then we have that
    \begin{equation*}
     \begin{aligned}
        &\mathcal{L}_{\mathcal{M}}[h] = (\text{Id} - \Pi_{\mathcal{M}})[g]\\
         \Longrightarrow\qquad &(\text{Id} - \Pi_{\mathcal{M}})[h] = \mathcal{L}^{-1}_{\mathcal{M}}\Big[(\text{Id} - \Pi_{\mathcal{M}})[g]\Big] =: \mathcal{L}^{-1}_{\mathcal{M}}[g] 
    \end{aligned}
    \end{equation*}
    and $\ \mathcal{L}_{\mathcal{M}}\left[\mathcal{L}^{-1}_{\mathcal{M}}[g]  \right] = (\text{Id} - \Pi_{\mathcal{M}})[g]$.
    
\end{enumerate}
For a Maxwellian $\mathcal{M} = \mathcal{M}_{\varrho, \bm{U}, \vartheta}$, the orthogonal projection onto $\mathscr{I}$ can be written as 
\begin{multline}\label{eq:orthproj}
    \Pi_{\mathcal M} [g] := \frac{1}{\varrho}\abrac{1 ,\,\mathcal M\, g} 
    +
    \frac{\bm v-\bm U}{\varrho\vartheta}
    \cdot
    \abrac{ 
    (\bm v- \bm U),\,\mathcal M \, g} 
    + 
    \left( \frac{|\bm v- \bm U|^2}{2\vartheta} -\frac{3}{2} \right) 
    \frac{2}{3\varrho}
    \abrac{
    \left( \frac{|\bm v- \bm U|^2}{2\vartheta} -\frac{3}{2} \right) , \, \mathcal M\, g}
\end{multline}
where the following orthonormal basis of $\mathscr{I}$ has been used
$$\left\{ \frac{1}{\sqrt{\varrho}},\, \frac{v_1-U_1}{\sqrt{\varrho\vartheta}}, \, \frac{v_2-U_2}{\sqrt{\varrho\vartheta}}, \, \frac{v_3 - U_3}{\sqrt{\varrho\vartheta}},\, \sqrt{\frac{2}{3\varrho}}\left(\frac{|\bm{v}-\bm{U}|^2}{2\vartheta} - \frac{3}{2}\right)\right\} =: \{\bar{e}_0,\,\bar{e}_1,\, \bar{e}_2,\, \bar{e}_3,\, \bar{e}_4 \}$$

\subsection{Derivative of the Inverse Linearized Collision Operator}\label{subsec:derivLinCol}
For the closures beyond the Navier-Stokes-Fourier level, the derivative of the inverse of the linearized collision operator, i.e. $\partial_{x_i}\left[\mathcal{L}_{\mathcal{M}}^{-1}[g] \right]$, becomes important to their derivation. For linearized collision operators of the form \eqref{eq:LinColOpDef}, this derivative can be expressed in terms of $\mathcal{M}$ and $\mathcal{L}_{\mathcal{M}}^{-1}$ and $Q$.

\begin{theorem}\label{thm:LinOpDerivative}
    Given a linearized collision operator of the form $\mathcal{L}_{\mathcal{M}}[g] = \frac{2}{\mathcal{M}}Q(\mathcal{M},\, \mathcal{M}g)$ for a Maxwellian $\mathcal{M}$ and function $g$ that are differentiable in $\bm{x}$ ({without loss of generality}), we have that 
 \begin{align}\label{eq:InvLDerivative}
    \partial_{x_i} \mathcal{L}_{\mathcal{M}}^{-1}\left[g \right] &= \mathcal{L}_{\mathcal{M}}^{-1}\left[\partial_{x_i}g \right]\, -\, \mathcal{L}_{\mathcal{M}}^{-1}\left[g \right]\, \partial_{x_i}\ln\mathcal{M}  + \mathcal{X}_{\mathcal{M}}\!\big[\dx{i}\!\ln\mathcal{M},\, g\big]
\end{align}
where $\mathcal{X}_{\mathcal{M}}$ is a bilinear operator defined as 
\begin{align}\label{eq:XtraTermDeriv}
    \mathcal{X}_{\mathcal{M}}[h, g] = \mathcal{L}^{-1}_{\mathcal{M}}\left[h\,(\textrm{Id}-\Pi_{\mathcal{M}})[g] - \frac{2}{\mathcal{M}}Q\left(\mathcal{M}h, \mathcal{M}\,\mathcal{L}_{\mathcal{M}}^{-1}[g]\right)\right]
\end{align}
\end{theorem}
\noindent 
An immediate consequence of Theorem \ref{thm:LinOpDerivative} is the following, more useful identity 
\begin{corollary}\label{cor:LinOpDerivative}
    \begin{align}\label{eq:DerivativeSwitchIdentity}
    \frac{1}{\mathcal{M}} \partial_{x_k}\left[\mathcal{M}\, \mathcal{L}_{\mathcal{M}}^{-1}\left[g \right] \right] &= \partial_{x_k}\mathcal{L}_{\mathcal{M}}^{-1}\left[g \right]+ \mathcal{L}_{\mathcal{M}}^{-1}\left[g \right]\,\partial_{x_k}\!\ln\mathcal{M}\nonumber  \\
    &= \mathcal{L}_{\mathcal{M}}^{-1}\left[\partial_{x_k}g \right]+\mathcal{X}_{\mathcal{M}}\big[\dx{k}\!\ln\mathcal{M},\, g \big]
\end{align}
\end{corollary}

In order to prove Theorem \ref{thm:LinOpDerivative}, we will need several lemmas.
\begin{lemma}\label{lem:projRange}
    The range of the linear operator $\partial_{x_i}\Pi_{\mathcal{M}}$ is in the space of collision invariants $\mathscr{I}$. 
\end{lemma}
\begin{proof}
    The projection onto the space of collision invariants is given by
    $$\Pi_{\mathcal{M}}[g] = \sum_{n=0}^4\bar{e}_n\abrac{\bar{e}_{n},\,\mathcal{M} g} $$ where $\bar{e}_i$ are the orthonormal collision invariants defined earlier. Taking the derivative yields
    $$\partial_{x_i}\Pi_{\mathcal{M}}[g] = \sum_{n=0}^4 \partial_{x_i}[\bar{e}_n] \abrac{\bar{e}_n, \,\mathcal{M} g } + \bar{e}_n\, \partial_{x_i}\!\left[\abrac{\bar{e}_n,\,\mathcal{M}\,g}\right]$$
and we observe that the above term remains a linear combination of collision invariants.   
\end{proof}

\begin{lemma}\label{lem:DLMterm}
    Given the linearized operator 
    $$\LM[g] = \int_{\mathbb{R}^3\times\mathbb{S}^2}\!\mathcal{M}_* (g_*' + g' - g - g_*) \mathfrak{B}(\bm{\hat{\omega}, \bm{v}, \bm{v_*}}) d\bm{\hat{\omega}}d\bm{v_*} = \mathtt{L}[\mathcal{M}, g]$$
    The derivative of the linearized operator takes the form 
    \begin{align}\label{eq:1stStepDeriv}
        \dx{i}\LM[g] = \LM[\dx{i}g] + \mathtt{L}[\mathcal{M}\ln\mathcal{M}, g]
    \end{align}
\end{lemma}
\begin{proof}
    By the product rule,
\begin{align*}
    \dx{i}\LM[g] &= \left.\frac{d}{d\eta}\mathtt{L}\left[{\mathcal{M}+\eta\,\partial_{x_i}\mathcal{M},\, g}\right] \right|_{\eta=0} +  \left.\frac{d}{d\eta}\mathtt{L}\left[{\mathcal{M},\, g + \eta\, \partial_{x_i} g}\right] \right|_{\eta=0}
    \\
    &= \mathtt{L}[\dx{i}\mathcal{M}, g] + \mathtt{L}[\mathcal{M}, \dx{i}g]
    \\
    &= \mathtt{L}[\mathcal{M}\dx{i}\!\ln\mathcal{M},\, g] + \LM[\dx{i}g]
\end{align*}  
\end{proof}

\begin{lemma}\label{lem:LMlnMterm}
The following identity holds
    \begin{align*}
        \mathtt{L}[\mathcal{M}\dx{i}\!\ln\mathcal{M},\,g] = \LM[g \,\dx{i}\!\ln\mathcal{M}] -\LM[g]\, \dx{i}\!\ln\mathcal{M}\, + \frac{2}{\mathcal{M}}Q\big( \, \mathcal{M}\,\dx{i}\!\ln\mathcal{M},\, \mathcal{M}\,g \big)
    \end{align*}
\end{lemma}
\begin{proof}
    Recall that because $\dx{i}\!\ln\mathcal{M}$ is a collision invariant, we have the conservation relation $$\dx{i}\!\ln\mathcal{M}' +\dx{i}\!\ln\mathcal{M}'_*- \dx{i}\!\ln\mathcal{M} - \dx{i}\!\ln\mathcal{M}_* = 0$$ 
Thus
\begin{align*}
     \mathtt{L}[\mathcal{M}\dx{i}\!\ln\mathcal{M},\,g] &= \int_{\mathbb{R}^3\times\mathbb{S}^2}\!\mathcal{M}_*\dx{i}\!\ln\mathcal{M}_* ({\color{olive} g_*' + g'} - g - {\color{red} g_*}) \mathfrak{B} d\bm{\hat{\omega}}d\bm{v_*}
     \\ &\hspace{0.5cm}
+ \int_{\mathbb{R}^3\times\mathbb{S}^2}\!\mathcal{M}_*\,g\, ({\color{blue} \dx{i}\!\ln\mathcal{M}} -{\color{red} \dx{i}\!\ln\mathcal{M}} ) \mathfrak{B}d\bm{\hat{\omega}}d\bm{v_*} 
   \\ &\hspace{0.5cm}
+ \int_{\mathbb{R}^3\times\mathbb{S}^2}\!\mathcal{M}_*\,g_*'\, ({\color{red} \dx{i}\!\ln\mathcal{M}_*'} + \dx{i}\!\ln\mathcal{M}' {\color{blue} - \dx{i}\!\ln\mathcal{M}} {\color{olive} - \dx{i}\!\ln\mathcal{M}_*}) \mathfrak{B}d\bm{\hat{\omega}}d\bm{v_*} 
   \\ &\hspace{0.5cm}
+ \int_{\mathbb{R}^3\times\mathbb{S}^2}\!\mathcal{M}_*\,g'\, (\dx{i}\!\ln\mathcal{M}_*' + {\color{red} \dx{i}\!\ln\mathcal{M}'} {\color{blue} - \dx{i}\!\ln\mathcal{M}} {\color{olive} - \dx{i}\!\ln\mathcal{M}_*}) \mathfrak{B}d\bm{\hat{\omega}}d\bm{v_*} 
\\&\hspace{0.5cm}
+ \int_{\mathbb{R}^3\times\mathbb{S}^2}\!\mathcal{M}_*\,g_*\, ({\color{blue} \dx{i}\!\ln\mathcal{M} } - \dx{i}\!\ln\mathcal{M}) \mathfrak{B}d\bm{\hat{\omega}}d\bm{v_*}
\end{align*}
Grouping the {\color{red} red} and {\color{blue} blue} terms together and canceling out the {\color{olive} olive} terms gives 
\begin{align*}
     \mathtt{L}[\mathcal{M}\dx{i}\!\ln\mathcal{M},\,g] &= {\color{red} \LM[g\, \dx{i}\!\ln\mathcal{M}]} - {\color{blue} \dx{i}\!\ln\mathcal{M}\,\LM[g]}
     \\ &\hspace{0.5cm}
     + \int_{\mathbb{R}^3\times\mathbb{S}^2}\!\mathcal{M}_* (g'_*\dx{i}\!\ln\mathcal{M}' + g'\dx{i}\!\ln\mathcal{M}_*' - g\dx{i}\!\ln\mathcal{M}_* - g_*\dx{i}\!\ln\mathcal{M}  )\mathfrak{B}d\bm{\hat{\omega}}d\bm{v_*}
\end{align*}
Because $\mathcal{M}'_* \mathcal{M}' = \mathcal{M}_*\mathcal{M}$, we can multiply the last term by $\frac{\mathcal{M}}{\mathcal{M}}$ and regroup the constituents of the last term so that we have 
\begin{align*}
     \mathtt{L}[\mathcal{M}\dx{i}\!\ln\mathcal{M},\,g] &= {\color{red} \LM[g\, \dx{i}\!\ln\mathcal{M}]} - {\color{blue} \dx{i}\!\ln\mathcal{M}\,\LM[g]}
     \\ &\hspace{0.5cm}
     + \frac{1}{\mathcal{M}}\int_{\mathbb{R}^3\times\mathbb{S}^2}\!\Big((g_*'\mathcal{M}_*')\mathcal{M}'\dx{i}\!\ln\mathcal{M}' + (g'\mathcal{M}')\mathcal{M}_*'\dx{i}\!\ln\mathcal{M}_*'
     \\ &\hspace{3.5cm}
     -(g_*\mathcal{M}_*)\mathcal{M}\dx{i}\!\ln\mathcal{M} - (g\mathcal{M})\mathcal{M}_*\dx{i}\!\ln\mathcal{M}_*\Big) \mathfrak{B}d\bm{\hat{\omega}}d\bm{v_*}
     \\
     &= { \LM[g\, \dx{i}\!\ln\mathcal{M}]} - { \dx{i}\!\ln\mathcal{M}\,\LM[g]} +\frac{2}{\mathcal{M}}Q\big(\mathcal{M}\,g,\, \mathcal{M}\,\dx{i}\!\ln\mathcal{M} \big)
\end{align*}
\end{proof}

\begin{lemma}\label{lem:colInvDeriv}
    \begin{align*}
        \Pi_{\mathcal{M}}\big[\dx{i}\iLM[g]\big] = - \Pi_{\mathcal{M}}\big[ \iLM[g]\,\dx{i}\!\ln\mathcal{M}\big]
    \end{align*}
\end{lemma}
\begin{proof}
     Note that for any function $\bar{h}(t, \bm{x}, \bm{v})\in \mathscr{I}$,
     
\begin{align*}
    \left\langle \bar{h},\, \mathcal{M}\, \partial_{x_i} \mathcal{L}_{\mathcal{M}}^{-1}\left[g \right] \right\rangle &=\partial_{x_i} \overbrace{\left\langle \bar{h} , \mathcal{M} \, \mathcal{L}_{\mathcal{M}}^{-1}\left[g \right]\right\rangle}^{=\,0} - \left\langle\frac{1}{\mathcal{M}} \partial_{x_i}\big[\bar{h}\,\mathcal{M}\big] , \mathcal{M} \, \mathcal{L}_{\mathcal{M}}^{-1}\left[g \right]\right\rangle\nonumber\\
    &= -\overbrace{\left\langle \partial_{x_i}\bar{h} ,\, \mathcal{M} \, \mathcal{L}_{\mathcal{M}}^{-1}\left[g \right]\right\rangle}^{=\,0} - \left\langle \bar{h}\, \partial_{x_i}\ln\mathcal{M} ,\, \mathcal{M} \, \mathcal{L}_{\mathcal{M}}^{-1}\left[g \right]\right\rangle\nonumber\\
    &= - \left\langle \bar{h} ,\, \mathcal{M} \, \mathcal{L}_{\mathcal{M}}^{-1}\left[g \right]\, \partial_{x_i}\ln\mathcal{M}\right\rangle
\end{align*}
The above identity will, in particular, be true for the orthonormal basis $\bar{e}_\beta$. Thus 
    \begin{align*}
        \Pi_{\mathcal{M}}\big[\dx{i}\iLM[g]\big] = - \Pi_{\mathcal{M}}\big[ \iLM[g]\,\dx{i}\!\ln\mathcal{M}\big]
    \end{align*}
    (Note that the same identity and proof will also hold for the linearized collision operator $\LM$.)
\end{proof}
We can now proceed with the main proof.
\begin{proof}[Proof of Theorem \ref{thm:LinOpDerivative}]
    \begin{align}\label{eqCalc:DiLMstart}
        \dx{i}[\LM[\iLM[g]]] &= \dx{i}\big[(\Id -\Pi_{\mathcal{M}})[g] \big] = \dx{i}g - \dx{i}\Pi_{\mathcal{M}}[g]
    \end{align}
Applying Lemmas~\ref{lem:DLMterm}~and~\ref{lem:LMlnMterm} to the left-hand side term gives
\begin{align*}
&\LM[\dx{i}\iLM[g]]+\LM[\iLM[g]\dx{i}\!\ln\mathcal{M}] 
        - \dx{i}\!\ln\mathcal{M}(\Id - \Pi_{\mathcal{M}})[g]+ \frac{2}{\mathcal{M}}Q\big(\mathcal{M}\iLM[g], \, \mathcal{M}\dx{i}\!\ln\mathcal{M} \big)
\end{align*}
Equation \eqref{eqCalc:DiLMstart} can then be rearranged to get
\begin{align*}
    \LM\left[ \dx{i}\iLM[g] +\iLM[g]\dx{i}\!\ln\mathcal{M}\right] &= \dx{i}g - \dx{i}\Pi_{\mathcal{M}}[g] \\ &\hspace{0.35cm}
    +\dx{i}\!\ln\mathcal{M}(\Id - \Pi_{\mathcal{M}})[g]\!-\! \frac{2}{\mathcal{M}}Q\big(\mathcal{M}\iLM[g],  \mathcal{M}\dx{i}\!\ln\mathcal{M} \!\big)
    \\
    (\Id - \Pi_{\mathcal{M}}\!)\!\!\left[ \dx{i}\iLM[g] +\iLM[g]\dx{i}\!\ln\mathcal{M}\right] \!&=\! \iLM[\dx{i}g] + \mathcal{X}_{\mathcal{M}}[\dx{i}\!\ln\mathcal{M},\,g]
\end{align*}
In the second line, we used Lemma~\ref{lem:projRange} and the fact the kernel of $\iLM$ is the space of collision invariants. Finally, applying the identity in Lemma~\ref{lem:colInvDeriv} gives
\begin{align*}
    \partial_{x_i} \mathcal{L}_{\mathcal{M}}^{-1}\left[g \right] &= \mathcal{L}_{\mathcal{M}}^{-1}\left[\partial_{x_i}g \right]\, -\, \mathcal{L}_{\mathcal{M}}^{-1}\left[g \right]\, \partial_{x_i}\ln\mathcal{M}  + \mathcal{X}_{\mathcal{M}}[\dx{i}\ln\mathcal{M},\, g]
\end{align*}
\end{proof}
\begin{remark}
   Identities of the form \eqref{eq:InvLDerivative} and \eqref{eq:DerivativeSwitchIdentity} will hold for other classes of collision operators such as the BGK collision operator, however the operator $\mathcal{X}_{\mathcal{M}}$ is no longer guaranteed to be of the form \eqref{eq:XtraTermDeriv} because \eqref{eq:1stStepDeriv} no longer applies. Lemma~\ref{lem:colInvDeriv} would still be true because it only relies on self-adjointness and collision invariance of the linearized operator.
\end{remark}

\section{The Variational Multiscale Moment Method}\label{sec:VMS}
To derive closed conservation equations from the Boltzmann equation, we first express the distribution $F$ as a perturbation of its self-consistent Maxwellian, i.e., $F = \mu(F)\,(1 +\epsilon f)$. The perturbation $f$ will belong in the orthogonal complement $\mathscr{I}^{\perp_{\mu(F)}}$ because, by construction, $\mu(F)$ is such that for any $\bar{m} \in \mathscr{I}$, $$\abrac{\bar{m},\,\mu(F) (1 + \epsilon f)} = \abrac{\bar{m},\, \mu(F)}$$
Substituting this ansatz into \eqref{eq:BoltzmannEqn} yields
\begin{equation}\label{eq:HydroBoltzmannEqn}
   (\text{St}\,\partial_t + v_i\partial_{x_i})[\mu + \epsilon \mu f] = \mu \mathcal{L}_{\mu}[f] + {\epsilon\,}Q(\mu f,\, \mu f)  
\end{equation}
where $\mu = \mu(F)$ for notational brevity. The conservation equations \eqref{eq:macroconslaws} arise from testing \eqref{eq:HydroBoltzmannEqn} with an appropriate collision invariant $\bar{m} \in \mathscr{I}$
\begin{equation}
\begin{aligned}\label{eq:CoarseScaleNonLinear}
    \abrac{\bar{m}, \text{St}\,\partial_{t}\mu} + \abrac{\bar{m}\,v_i,\,\partial_{x_i}\mu} + \epsilon \abrac{\bar{m}\,v_i,\,\partial_{x_i}[\mu f]} &= 0
    \\
     \Leftrightarrow \St\abrac{\bar{m},\, \mu\,\, \partial_{t}\!\ln \mu} + \abrac{\bar{m}v_i,\, \mu\, \dx{i}\!\ln\mu} +\epsilon \abrac{\bar{m}, \vdx{i}[\mu\, f] }&= 0
\end{aligned}
\end{equation}
In the language of the Variational Multiscale method, the above equation is referred to as the \textit{coarse-scale equation}. The deviatoric stress $\sigma_{ij}$ and heat flux $q_i$ that we seek closure relations for arise from the term $\abrac{\bar{m}\,v_i,\,\partial_{x_i}[\mu f]}$. Deriving closures thus amount to substituting into \eqref{eq:CoarseScaleNonLinear} some approximation or model of $f$ in terms of $\mu$. A starting point for obtaining these \textit{fine-scale approximations}, denoted $f_{(n)}(\mu)$, is found by rearranging Equation \eqref{eq:HydroBoltzmannEqn} to give the following expression for $f$
\begin{equation}\label{eq:FineScaleNonLinear}
    f = \mathcal{L}_{\mu}^{-1}\left[v_i\,\partial_{x_i}\!\ln\mu + \frac{1}{\mu}(\text{St}\,\partial_t + v_j\partial_{x_j})[\mu f] - \frac{1}{\mu}Q(\mu f,\, \mu f) \right]
\end{equation}
By coming up with approximations of the right-hand side of \eqref{eq:FineScaleNonLinear} in terms of $\mu$, we can close the coarse scale equation. With $\bar{m} = \ln\mu$, the coarse-scale equation \eqref{eq:CoarseScaleNonLinear} can be rewritten as 
\begin{equation}\label{eq:CoarseScaleNonLinearEntropy}
    \text{St}\,\partial_t\abrac{\ln\mu - 1, \mu }+\partial_{x_i}\abrac{\ln\mu - 1, v_i\,\mu} + \epsilon \partial_{x_i}\abrac{v_i\ln\mu,\, \mu f} = \epsilon \abrac{v_k\partial_{x_k}\!\ln\mu\,,\,\mu f}
\end{equation}
An important design criterion for the fine-scale approximations $f_{(n)}(\mu)$ is \textit{coarse-scale entropy stability} where $f_{(n)}(\mu)$ is such that under situations where boundary contributions can be ignored (for example, periodic boundary conditions), Equation \eqref{eq:CoarseScaleNonLinearEntropy} can be written as
\begin{equation}\label{eq:NonLinearCoarseEntropyStability}
   \text{St}\,\frac{d}{dt} \int_\Omega \abrac{\ln\mu - 1, \mu }\, d\bm{x} = \epsilon\int_{\Omega}\abrac{v_k\partial_{x_k}\!\ln\mu\,,\,\mu f_{(n)}(\mu)} \, d\bm{x}\le 0
\end{equation}
Comparing this to the entropy dissipation inequality \eqref{eq:entdissip}
\begin{equation}\label{eq:entdissipSpatial}
   \text{St}\,\frac{d}{dt} \int_\Omega \abrac{\ln F - 1, F }\, d\bm{x} = \frac{1}{\epsilon}\int_{\Omega}\abrac{\ln(F)\,,\,Q(F,\,F)} \, d\bm{x}\le 0
\end{equation}
we see that requiring coarse-scale entropy stability for the fine-scale approximation amounts to requiring that closed conservation equations possess a "discrete"\footnote{The terminology "discrete" is used to indicate that $\ln\mu \in \mathscr{I}$ is a finite-dimensional  approximation of $\ln F$ in $\bm{v}$.} entropy dissipation inequality that mirrors that of the Boltzmann equation.

\begin{remark}
    The coarse-scale entropy $\abrac{\ln\mu - 1,\, \mu}$ and coarse-scale entropy flux $\abrac{\ln\mu - 1, v_i\,\mu}$ in terms of the conservation variables $\rho,\,\bm{u}$ and $\theta$ will be
    \begin{align}
         \abrac{\ln\mu - 1,\, \mu} &= \rho \left(\ln\left(\frac{\rho}{(2\pi\theta)^{\frac{3}{2}}} \right) - \frac{5}{2}\right)\quad =: H(t, \bm{x}) \label{eq:MacroEntropy}\\
        \abrac{\ln\mu - 1,\, v_i\, \mu} &= \rho \left(\ln\left(\frac{\rho}{(2\pi\theta)^{\frac{3}{2}}} \right) - \frac{5}{2}\right)\,u_i \quad= H\,u_i \label{eq:MacroEntropyFlux} 
    \end{align}
   {Observe that the Sackur-Tetrode equation for the entropy and \eqref{eq:MacroEntropy} are related by
\begin{equation}\label{eq:Sackur-Tetrode}
   \frac{S}{k_B N} = \ln \parent{\frac{V}{N}\parent{\frac{4\pi m U}{3h^2 N}}^{\frac{3}{2}}} + \frac{5}{2} = \ln\parent{\frac{m^4}{h^3}}-\frac{1}{\rho} \abrac{\ln\mu - 1,\, \mu}  
   \end{equation}
    where $h$ is Planck's constant, $N = \rho V/m$ is the number of particles in the volume $V$ and $U = \frac{3}{2}mN \theta$. } 
\end{remark}

In this work, we shall focus on the following fine-scale approximations that satisfy the coarse-scale entropy dissipation relation:
\begin{enumerate}
    \item \textit{Euler}: 
    \begin{equation}
        \label{eq:EulerfinescaleNL} f_{(0)} = 0
    \end{equation}
    With this approximation, \eqref{eq:NonLinearCoarseEntropyStability} becomes 
     $$ \text{St}\,\frac{d}{dt} \int_\Omega \abrac{\ln\mu - 1, \mu }\, d\bm{x} =  0$$
    
    \item \textit{Navier-Stokes-Fourier (NSF)}:
    \begin{equation}\label{eq:NSFfinescaleNL}
        f_{(1)} = \mathcal{L}_{\mu}^{-1}\left[ v_i \partial_{x_i}\!\ln(\mu) \right]
    \end{equation}
    This approximation \eqref{eq:NSFfinescaleNL} is obtained by treating the fine-scale equation \eqref{eq:FineScaleNonLinear} as a fixed point problem initialized with $f_{(0)} = 0$, i.e.
    \begin{equation}\label{eq:Chapman-Enskog}
     f_{(1)} = \mathcal{L}_{\mu}^{-1}\left[v_i\,\partial_{x_i}\!\ln\mu + \frac{1}{\mu}(\text{St}\,\partial_t + v_j\partial_{x_j})[\mu f_{(0)}] - \frac{1}{\mu}Q(\mu f_{(0)},\, \mu f_{(0)}) \right]
       \end{equation}
    Substituting the approximation \eqref{eq:NSFfinescaleNL} into the right-hand side of \eqref{eq:NonLinearCoarseEntropyStability} gives
    $$ \text{St}\,\frac{d}{dt} \int_\Omega \abrac{\ln\mu - 1, \mu }\, d\bm{x} = \epsilon\int_{\Omega}\abrac{v_k\partial_{x_k}\!\ln\mu\,,\,\mu\, \mathcal{L}_{\mu}^{-1}\left[v_i\partial_{x_i}\ln(\mu) \right]} \, d\bm{x}\le 0$$
    where the inequality is a result of the negative semi-definiteness of the linearized collision operator.
    
    It can be shown that the above fixed point iteration \eqref{eq:Chapman-Enskog} is equivalent to the Chapman-Enskog expansion \cite{chapman1990mathematical, Enskog1917}. The next fine-scale approximation, $f_{(2)}$, that results from the procedure \eqref{eq:Chapman-Enskog} yields the Burnett equations which are, unfortunately, widely known to be unstable \cite{Agarwal2001, bobylev1982chapman, Comeaux1995, GARCIACOLIN2008}.
We find that this instability manifests as a failure to exhibit coarse-scale entropy stability \eqref{eq:NonLinearCoarseEntropyStability}. This motivated the creation of a fine-scale approximation that improves upon the Navier-Stokes-Fourier closure while exhibiting coarse-scale entropy stability.
    \item \textit{Entropy Stable Extension to Navier-Stokes-Fourier (ESE)}:  In \cite{Baidoo2026}, we introduced the Entropy Stable Extension to the Navier-Stokes-Fourier equations which were obtained with a modified fixed-point problem derived from the fine-scale equation \eqref{eq:FineScaleNonLinear} under the assumption that the Strouhal number is $O(\epsilon^n)$ for $n\ge 1$. The resulting fine-scale approximation is 
 \begin{align}\label{eq:ESEfinescaleNL}
        {f}_{(2^E)} &= {f}_{(1)} -\epsilon^2 \mathcal{L}_\mu^{-1}v_k\partial_{x_k}\left[ \mathcal{L}_\mu^{-1}\left[ \frac{1}{\mu} v_j\partial_{x_j}\left[\mu \,\mathcal{L}_\mu^{-1}[v_i\partial_{x_i}\!\ln\left(\mu \right)]\right]\right]\right]
    \end{align}
 Substituting the above approximation into the right-hand side of \eqref{eq:NonLinearCoarseEntropyStability} and integrating by parts while taking advantage of the self-adjointness of the linearized operator yields
\begin{multline*}
     \text{St}\,\frac{d}{dt} \int_\Omega \abrac{\ln\mu - 1, \mu }\, d\bm{x} = \epsilon\int_{\Omega}\abrac{v_k\partial_{x_k}\!\ln\mu\,,\,\mu\, \mathcal{L}_{\mu}^{-1}\left[v_i\partial_{x_i}\ln(\mu) \right]} \, d\bm{x} \\
     + \epsilon^3 \int_{\Omega} \abrac{\frac{1}{\mu}v_k \partial_{x_k}\left[\mu{f}_{(1)}\right], \mu\,   \mathcal{L}_\mu^{-1}\left[\frac{1}{\mu} v_j\partial_{x_j}\left[\mu {f}_{(1)}\right]\right]}\, d\bm{x} \le 0
\end{multline*}
which shows that the resulting closure is entropy stable.  
\end{enumerate}

\section{Entropy Variable Description of Conservation Equations derived from the Boltzmann Equation}\label{sec:EntropyVariables}
Insight into how the structure of the ESE equations leads to entropy dissipation is best gained by working in entropy variables:
\begin{equation}\label{eq:EntropyVariablesDefinition}
   \bm{V}(t, \bm{x}) := 
    \begin{pmatrix}
        \ln\!\left(\frac{\rho}{(2\pi\theta)^{\frac{3}{2}}}\right) - \frac{|\bm{u}|^2}{2\theta}
        \\
        \frac{u_1}{\theta}
        \\
        \frac{u_2}{\theta}
        \\
        \frac{u_3}{\theta}
        \\
        -\frac{1}{\theta}
    \end{pmatrix}
\end{equation}
Introducing the basis vector $\bm{\bar{b}}(\bm{v}) = (1, v_1, v_2, v_3, |\bm{v}|^2/2)^{\mathsf{T}}$, we can immediately observe that $ \ln\mu = {V}_\alpha(t, \bm{x})\, {\bar{b}_\alpha}(\bm{v})$, where we employ a repeated lowercase Greek alphabet to indicate summation over entropy variable indices (i.e. $\alpha$ runs over $0, 1,\dots,4$). 

Starting with the Euler equations, observe that 
\begin{equation}\label{eq:EulerEntropyVar}
    \begin{aligned}
\St\abrac{\bar{b}_\alpha,\, \mu\,\, \partial_{t}\!\ln \mu} + \abrac{\bar{b}_\alpha v_i,\, \mu\, \dx{i}\!\ln\mu} &= 0 \\
\Longleftrightarrow \ \St\abrac{\bar{b}_\alpha, \mu\,  \bar{b}_\beta} \partial_t V_\beta + \abrac{\bar{b}_\alpha v_i, \mu\, \bar{b}_\beta} \dx{i}V_\beta &= 0
    \end{aligned}
\end{equation}
Because the matrix $\bm{\mathrm{A}}^0_{\alpha\beta}(\bm{V}):= \St\abrac{\bar{b}_\alpha, \mu\,  \bar{b}_\beta}$ is symmetric positive-definite and the matrix $\bm{\mathrm{A}}^i_{\alpha\beta}(\bm{V}):= \abrac{\bar{b}_\alpha v_i, \mu\, \bar{b}_\beta}$ is symmetric for each $i$, we have just shown that the Euler equations in entropy variables take the form of a symmetric hyperbolic (or Friedrichs') system. 
Note that the dependence of the matrices on $\bm{V}$ comes from the Maxwellian $\mu = e^{\bm{V}\cdot\bm{\bar{b}}}$.

For the Navier-Stokes-Fourier equations, we have  
\begin{equation}\label{eq:NSFentropyVar}
    \begin{aligned}
      \St\abrac{\bar{b}_\alpha,\, \mu\,\, \partial_{t}\!\ln \mu} + \abrac{\bar{b}_\alpha v_i,\, \mu\, \dx{i}\!\ln\mu}+ \epsilon \abrac{\bar{b}_\alpha, \vdx{j}[\mu\, \mathcal{L}_{\mu}^{-1}\left[ v_k \partial_{x_k}\!\ln\mu \right]] }&= 0
        \\
        \Longleftrightarrow 
        \bm{\mathrm{A}}^0_{\alpha\beta}(\bm{V})\, \partial_t\! V_\beta + \bm{\mathrm{A}}^i_{\alpha\beta}(\bm{V})\,\dx{i}\!V_\beta + \epsilon\, \dx{j}\!\sqbrac{\abrac{\bar{b}_\alpha v_j, \, \mu \, \iLmu\big[\bar{b}_\beta v_k \big]} \dx{k}\!V_\beta} &= 0
    \end{aligned}
\end{equation}
Let $\bm{\mathrm{K}}^{jk}_{\alpha\beta}(\bm{V}) := - \epsilon \abrac{\bar{b}_\alpha v_j, \, \mu \, \iLmu\big[v_k \bar{b}_\beta\big]}$ and observe that due to the self-adjointness and negative semi-definiteness of the linearized collision operator, for any unit vector $\bm{\hat{\omega}}$, the matrix  $\bm{\mathrm{K}^{\hat{\omega}}}_{\alpha \beta}:= \hat{\omega}_j \bm{\mathrm{K}}^{jk}_{\alpha\beta} \hat{\omega}_k$ will be symmetric and positive semi-definite. This defines a symmetric hyperbolic-parabolic system of equations in the sense described in \cite{Kawashima1988}. It also corresponds to the symmetric advective-diffusive system as described in \cite{hughes1986I}. 

Finally the ESE contributions:
\begin{equation*}
    \begin{aligned}
       &- \abrac{\bar{b}_\alpha, \vdx{l}\mathcal{L}_\mu^{-1}v_k\partial_{x_k} \!\! \left[ \mathcal{L}_\mu^{-1} \!\! \left[ \frac{1}{\mu} v_j\partial_{x_j} \!\! \left[\mu \,\mathcal{L}_\mu^{-1}[v_i\partial_{x_i} \! \ln\mu ]\right]\right]\right]}
       \\ &\hspace{.5cm}
       = -\dx{l} \!\! \abrac{\iLmu\!\!\sqbrac{v_l \bar{b}_\alpha} \! , \mu \, \vdx{k} \! \iLmu \!\! \sqbrac{\frac{1}{\mu} \vdx{j} \!\! \sqbrac{\mu \iLmu[\vdx{i} \! \ln\mu]} }}
       \\&\hspace{.5cm}
       = -\dx{l}\dx{k} \!\! \abrac{v_k \iLmu \! \sqbrac{v_l\bar{b}_\alpha}, \mu \, \iLmu \!\! \sqbrac{\frac{1}{\mu} \vdx{j} \!\! \sqbrac{\mu \iLmu[\vdx{i} \! \ln\mu]}}}
       \\ &\hspace{1.5cm}
       + \dx{l} \! \abrac{\frac{1}{\mu}\vdx{k} \! \sqbrac{\mu \iLmu \! \sqbrac{v_l \bar{b}_\alpha}}, \mu \, \iLmu \!\! \sqbrac{\frac{1}{\mu} \vdx{j} \!\! \sqbrac{\mu \iLmu[\vdx{i} \! \ln\mu]}}}
    \end{aligned}
\end{equation*}

From Corollary~\ref{cor:LinOpDerivative}, we have 
\begin{equation*}
\begin{aligned}
    \frac{1}{\mu}\vdx{j}\!\sqbrac{\mu \iLmu[\vdx{i}\ln\mu]} &= v_j\iLmu\!\!\sqbrac{v_i\bar{b}_\beta}\dx{i}\!\dx{j}\!V_\beta + v_j\mathcal{X}_\mu\!\big[\bar{b}_\gamma,\, v_i\bar{b}_\beta \big]\dx{j}\!V_\gamma\,\dx{i}\!V_\beta 
    \\
    \frac{1}{\mu}\vdx{k} \!\! \sqbrac{\mu \iLmu[v_l \bar{b}_\alpha]} &= v_k \mathcal{X}_\mu \!\! \sqbrac{\,\bar{b}_\eta , v_l \bar{b}_\alpha}\, \dx{k}\!V_\eta
\end{aligned}  
\end{equation*}
and thus the ESE contribution takes the form
{\small
\begin{equation}\label{eq:ESEentropyVarContribution}
    \dx{l}\dx{k} \!\! \sqbrac{\Ltens{k}{l}{j}{i}{\alpha}{\beta}\, \dx{i}\dx{j}\!V_\beta } - \dx{l}\dx{k} \!\! \sqbrac{\Xtens{k}{l}{j}{i}{\alpha}{\gamma}{\beta}\,\dx{j}\!V_\gamma \,\dx{i}\!V_\beta} + \dx{l}\!\!\sqbrac{\Xtens{k}{i}{j}{l}{\beta}{\gamma}{\alpha}\,\dx{j}\!V_\gamma \, \dx{k}\dx{i}\!V_{\beta}} - \dx{l}\!\! \sqbrac{\Ytens{k}{l}{j}{i}{\eta}{\alpha}{\gamma}{\beta}\,\dx{k}\!V_{\eta}\, \dx{j}\!V_{\gamma}\,\dx{i}\!V_{\beta}}
\end{equation}
}
where 
\begin{subequations}\label{eq:ESEtenDef}
  \begin{align}
   \Ltens{k}{l}{j}{i}{\alpha}{\beta}(\bm{V}) &\coloneqq - \epsilon^3\abrac{v_k \iLmu \!\! \sqbrac{v_l \bar{b}_\alpha}, \mu\, \iLmu \!\! \sqbrac{ v_j \iLmu \!\! \sqbrac{v_i \bar{b}_\beta}}} \label{eq:LtensESE}
   \\
   \Xtens{k}{l}{j}{i}{\alpha}{\gamma}{\beta}(\bm{V}) &\coloneqq \ \epsilon^3 \abrac{v_k \iLmu\!\!\sqbrac{v_l \bar{b}_\alpha}, \mu\, \iLmu\!\!\sqbrac{v_j \mathcal{X}_\mu \!\! \sqbrac{\,\bar{b}_\gamma,\, v_i\bar{b}_\beta}}} \label{eq:XtensESE}
   \\ \label{eq:YtensESE}
   \Ytens{k}{l}{j}{i}{\eta}{\alpha}{\gamma}{\beta}(\bm{V}) &\coloneqq -\epsilon^3\abrac{v_k \mathcal{X}_\mu \!\! \sqbrac{\,\bar{b}_\eta,\, v_l\bar{b}_\alpha}, \mu\, \iLmu\!\!\sqbrac{v_j \mathcal{X}_\mu \!\! \sqbrac{\,\bar{b}_\gamma,\, v_i\bar{b}_\beta}}}
\end{align}  
\end{subequations}
Thus the ESE equations take the form of a fourth-order quasi-cubic system. By "quasi-cubic", we mean that the derivatives of $\bm{V}$ have at most a cubic non-linearity in \eqref{eq:ESEentropyVarContribution}. 

The quasi-linear approximation to the ESE equations
\begin{equation}\label{eq:qlESEeqnsEntropyVariableForm}
    \begin{aligned}
       \bm{\mathrm{A}}^0_{\alpha\beta}(\bm{V})\, \partial_t\! V_\beta + \bm{\mathrm{A}}^i_{\alpha\beta}(\bm{V})\,\dx{i}\!V_\beta - \dx{j} \!\! \sqbrac{\bm{\mathrm{K}}_{\alpha\beta}^{jk}(\bm{V})\  \dx{k}\!V_\beta} + \dx{l}\dx{k} \!\!\sqbrac{\Ltens{k}{l}{j}{i}{\alpha}{\beta}(\bm{V}) \ \dx{j}\dx{i}\!V_\beta} = 0
    \end{aligned}
\end{equation}
 deserves special attention. Notice that all the fourth-order derivative contributions are contained in the quasi-linear contribution $ \dx{l}\dx{k} \!\! \sqbrac{\Ltens{k}{l}{j}{i}{\alpha}{\beta}\, \dx{i}\dx{j}\!V_\beta }$ and that for any unit vector $\bm{\hat{\omega}}$, the matrix $\bm{\mathrm{L}^{\hat{\omega}}}_{\alpha\beta }:= \hat{\omega}_i \hat{\omega}_j \hat{\omega}_k \hat{\omega}_l  \Ltens{k}{l}{j}{i}{\alpha}{\beta}$ will be symmetric and positive semi-definite. Furthermore, the conservation equations that result from keeping just the quasi-linear contribution to the ESE equations remains entropy stable. This can be proved with the entropy variable formulation by observing that testing the coarse-scale equations \eqref{eq:CoarseScaleNonLinear} with $\bar{m} = \ln \mu$ and integrating in space and velocity is equivalent to testing the entropy variable form of the conservation equations with $V_\alpha$ and integrating in space. Under the same assumptions on spatial boundary terms as \eqref{eq:NonLinearCoarseEntropyStability}, this procedure applied to the quasi-linear ESE equations \eqref{eq:qlESEeqnsEntropyVariableForm}, after integrating by parts where appropriate, becomes 
\begin{equation*}
    \begin{aligned}
       &\qquad \int_{\Omega} V_\alpha \bm{\mathrm{A}}_{\alpha\beta}^0\, \partial_t\! V_\beta\, d\bm{x} = -\int_{\Omega} \dx{j}\!V_\alpha \,\bm{\mathrm{K}}^{jk}_{\alpha\beta} \, \dx{k}V_\beta\,d\bm{x} - \int_{\Omega} \dx{l}\dx{k}\!V_\alpha \ \Ltens{k}{l}{j}{i}{\alpha}{\beta} \ \dx{j}\dx{i}\!V_\beta \,d\bm{x}\ \le 0 \\
       &\Longleftrightarrow \St\,\frac{d}{dt} \int_\Omega \abrac{\ln\mu - 1, \mu }\, d\bm{x}  =  \epsilon\int_{\Omega}\abrac{v_j\partial_{x_j}\!\ln\mu\,,\,\mu\, \mathcal{L}_{\mu}^{-1}\left[v_k\partial_{x_k}\ln\mu \right]} \, d\bm{x}
       \\ &\hspace{5.5cm}
       + \epsilon^3\int_{\Omega}\abrac{v_k \iLmu \!\! \sqbrac{\vdx{l}\dx{k}\!\ln\mu},\,\mu \, \iLmu \!\! \sqbrac{v_j\iLmu \!\! \sqbrac{\vdx{i}\dx{j}\ln\mu}}}\,d\bm{x} \ \le 0
    \end{aligned}
\end{equation*}

\begin{remark}
   Godunov \cite{Godunov1961}, Friedrichs and Lax \cite{friedrichs1971} and Mock \cite{Mock1980} cumulatively showed the equivalence between the existence of a symmetric hyperbolic form and the existence of an entropy for a system of equations. Building on work of Harten \cite{Harten1983} and Tadmor \cite{Tadmor1984_2}, Hughes et al. \cite{hughes1986I} showed that the entropy variable form of the Navier-Stokes-Fourier equations restricted the entropy for the system to at most an affine function of the Sackur-Tetrode entropy \eqref{eq:Sackur-Tetrode}. Kawashima and Shizuta \cite{Kawashima1984,Kawashima1988} proved, more generally, that the symmetrization of a class of second order non-linear systems that includes the Navier-Stokes-Fourier equations is equivalent to the existence of an entropy function.
   
   The above analysis of the ESE equations and its quasi-linear approximation suggests that a similar result linking symmetrization to existence of an entropy function exists for some class of fourth order non-linear equations.
\end{remark}

\section{The quasi-linear approximation to the ESE equations}\label{sec:quasilinearESE}
The deviatoric stress and heat flux due quasi-linear approximation to the ESE take the form
\begin{subequations}\label{eq:qlESEclosure}
\begin{align}
    &\sigma_{ij}^{(2^{SE})} = 2 \omega\, \tepsilonu{ij} - \dx{k}\!\mathfrak{S}_{ijk},\label{eq:SESEdevStress}\\
    &q_i^{(2^{SE})} = - \kappa\, \dx{i}\theta +\dx{k} \mathfrak{H}_{ik}+ \dx{k}\!u_j\,\mathfrak{S}_{ijk},\label{eq:SESEheatFlux}
\end{align}
\end{subequations}
which when substituted into \eqref{eq:macroconslaws} results in a closed system of conservation equations of the form 
\begin{subequations}
\label{eq:ESEmacroconslaws}
\begin{align}
\text{St }\partial_t\rho + \partial_{x_i} [\rho u_i]&=0 \label{eq:ESEmasslaw}\\
\text{St }\partial_t[\rho u_j] + \partial_{x_i} [\rho {u}_i{u}_j+ \rho \theta\delta_{ij} - 2\omega\, \tepsilon_{ij}(\bm{u}) + \partial_{x_k}\mathfrak{S}_{ijk}]&=0\label{eq:ESEmomentumlaw}\\
\text{St }\partial_t \!\! \left[\frac{1}{2}\rho|\bm u|^2+\frac{3}{2}\rho \theta\right] + \partial_{x_i} \! \Bigg[\!\! \left(\frac{1}{2}\rho|\bm u|^2 + \frac{3}{2} \rho\theta \right)\! u_i+ \rho\theta u_i- 2\omega\,\tepsilon_{ij}(\bm{u})\, u_j -\kappa \partial_{x_i}\theta
+ \partial_{x_k}\!\left[u_j\mathfrak{S}_{jik} + \mathfrak{H}_{ik} \right]\Bigg]&=0\label{eq:ESEenergylaw}
\end{align}
\end{subequations}

The tensors $\tepsilonu{ij}, \,\mathfrak{H}_{ij} \text{ and } \mathfrak{S}_{ijk}$ are given by
\begin{subequations}\label{eq:qESEClosureTensors}
\begin{align}
\label{eq:symmetricGradient}
\tepsilonu{ij} &= \left(\frac{\dx{i}u_j + \dx{j}u_i}{2} - \frac{\delta_{ij}}{3}\divu \right)
\end{align}
\begin{equation}
    \begin{aligned}
        \label{eq:ESEHeatTensor}
      \mathfrak{H}_{ij} &=\Lambda_1\,\theta^2\, \Delta\! V_4 \delta_{ij} + \Lambda_2\, \theta^2 \dx{i}\dx{j}\!V_4 + \mathcal{B}\,\theta^2\,u_j\uD[\dx{i}\!V_4] + 2\mathcal{D}\,\theta^2 \parent{ \uD\sqbrac{\frac{\tepsilonu{ij}}{\theta}} - \frac{1}{2}\dx{a}\!V_4\,\KrPhi_{ij}^{ab}\,\uD u_b} 
      \\&\hspace{.5cm}
      + 2\mathcal{P}\, \theta^2 \, u_j  \parent{\dx{n}\sqbrac{\frac{\tepsilonu{ni}}{\theta}} - \frac{1}{2}\dx{a}\!V_4\,\KrPhi_{ib}^{ac} \,\dx{b}\!u_c } 
    \end{aligned}
\end{equation}
\begin{align}
\label{eq:ESESpress}
\mathfrak{S}_{ijk} &=  2 {\alpha}_1\theta \left(\dx{k}\!\!\sqbrac{\frac{\tepsilon_{ij}(\bm{u})}{\theta}} +  \dx{i}\!\!\sqbrac{\frac{\tepsilon_{jk}(\bm{u})}{\theta}}+  \dx{j}\!\!\sqbrac{\frac{\tepsilon_{ik}(\bm{u})}{\theta}}  -\frac{2}{3}\delta_{ij}\dx{l}\!\!\sqbrac{\frac{\tepsilon_{lk}(\bm{u})}{\theta}} \right)\nonumber\\ \nonumber&\hspace{.35cm}
      +2\,{\alpha}_2 \theta \left(\delta_{ik}\,\dx{l}\!\!\sqbrac{\frac{\tepsilon_{lj}(\bm{u})}{\theta}} +\delta_{jk}\,\dx{l}\!\!\sqbrac{\frac{\tepsilon_{li}(\bm{u})}{\theta}} - \frac{2}{3}\delta_{ij}\,\dx{l}\!\!\sqbrac{\frac{\tepsilon_{lk}(\bm{u})}{\theta}}\right) 
      \\ \nonumber&\hspace{.5cm}
      -\alpha_1 \theta\! \parent{\dx{a}\!V_4\,\KrPhi_{ij}^{ab}\,\dx{k}\!u_b +\dx{a}\!V_4\,\KrPhi_{ij}^{ab}\,\dx{b}\!u_k + \dx{k}\!V_4\,\KrPhi_{ij}^{ab}\,\dx{a}\!u_b - \frac{2}{3}\dx{a}\!V_4\,\KrPhi_{ij}^{kb}\,\dx{b}\!u_a}
      \\
      \nonumber&\hspace{.5cm}
      -\alpha_2\theta \parent{\dx{a}\!V_4\,\KrPhi_{ij}^{bk}\,\dx{a}\!u_b + \dx{a}\!V_4\,\KrPhi_{ij}^{ak}\divu - \frac{2}{3}\dx{a}\!V_4\,\KrPhi_{ij}^{kb}\dx{b}\!u_a}
      \\ &\hspace{.5cm}\nonumber
            +2 \mathcal{A}\,\theta\, u_k \parent{\uD\!\!\sqbrac{\frac{\tepsilon_{ij}(\bm{u})}{\theta}} - \frac{1}{2} \dx{a}V_4 \,\KrPhi_{ij}^{ab}\,\uD u_b}
      \\ &\hspace{.5cm}
      + 2\mathcal{P}\theta^2\, \uD\!\left[\frac{\delta_{ik}\,\partial_{x_j}\!V_4 +\delta_{jk}\,\partial_{x_i}\!V_4}{2} - \frac{1}{3}\delta_{ij}\,\partial_{x_k}\!V_4  \right]  + 2\mathcal{D}\,\theta^2 u_k \left(\partial_{x_i}\partial_{x_j} - \frac{\delta_{ij}}{3}\Delta\right)\!\!\big[V_4\big]
\end{align}
\end{subequations}
where $V_4 = -1/\theta$, $\KrPhi_{ij}^{ab} = \delta_{ia} \delta_{jb} + \delta_{ib}\delta_{ja} - \frac{2}{3}\delta_{ij}\delta_{ab} $. In Appendix \ref{app:FluidComputeProcess}, we briefly describe how the terms \eqref{eq:qESEClosureTensors} are derived.
Table \ref{tab:nonLinearFluidParameters} gives the fluid parameters for Hard Spheres and Maxwell Molecules collision operators. The procedure by which the \eqref{eq:qESEClosureTensors} and the fluid parameters were calculated is detailed in \cite{Baidoo2026,Baidoo2025}. 
\begin{table}[ht]
    \centering
    {\tabulinesep=2mm
    \begin{tabu}{|m{6cm}|  m{7cm}| m{7cm} | }
    \hline
    \multicolumn{1}{|c|}{\textbf{Fluid Parameters}} & \multicolumn{1}{c|}{\textbf{Maxwell Molecules}} & \multicolumn{1}{c|}{\hspace{.2cm}\textbf{ Hard Spheres}\hspace{.4cm}}
    \\
    \hline
    \multicolumn{1}{|c|}{$ \tau$} &\multicolumn{1}{c|}{ $2\pi L \sqrt{\frac{2m}{V_0}}$} & \multicolumn{1}{c|}{$L  \sqrt{{\pi}/ ({8 \,\theta(t, \bm{x})})}$ } 
    \\   
     \hline
    \multicolumn{1}{|l|}{$\tilde{\omega}\coloneqq (\epsilon \tau \rho_0 \theta)^{-1}\, \omega$} &\multicolumn{1}{c|}{0.3440037} & \multicolumn{1}{c|}{1.270042 } 
    \\
    \hline
        \multicolumn{1}{|l|}{$\tilde{\kappa}\coloneqq\frac{2}{5}(\epsilon\tau \rho_0 \theta)^{-1}\,\kappa$} &\multicolumn{1}{c|}{0.5160055} & \multicolumn{1}{c|}{1.922284} 
    \\
     \hline
     \multicolumn{1}{|l|}{$\tilde{\mathcal{A}}\coloneqq\left(\epsilon^3\tau^3 \,{\rho_0^3 \theta}/{\rho^2}\right)^{-1}  \mathcal{A}$ } & \multicolumn{1}{c|}{0.04070889} & \multicolumn{1}{c|}{2.134497}
     \\ 
      \hline
     \multicolumn{1}{|l|}{$\tilde{\mathcal{B}}\coloneqq\frac{2}{5}\left(\epsilon^3\tau^3 \,{\rho_0^3 \theta}/{\rho^2}\right)^{-1}  \mathcal{B}$ } & \multicolumn{1}{c|}{0.1373925} & \multicolumn{1}{c|}{7.440228}
     \\
      \hline
     \multicolumn{1}{|l|}{$\tilde{\mathcal{D}}\coloneqq \left(\epsilon^3\tau^3 \,{\rho_0^3 \theta}/{\rho^2}\right)^{-1}  \mathcal{D}$ } & \multicolumn{1}{c|}{0.06106334} & \multicolumn{1}{c|}{1.946232}
     \\
      \hline
     \multicolumn{1}{|l|}{$\tilde{\mathcal{P}} \coloneqq \left(\epsilon^3\tau^3 \,{\rho_0^3 \theta}/{\rho^2}\right)^{-1} \mathcal{P}$ } & \multicolumn{1}{c|}{0.09159501} & \multicolumn{1}{c|}{3.856049}
     \\ 
      \hline
     \multicolumn{1}{|l|}{$\tilde{\alpha}_1 \coloneqq \left(\epsilon^3\tau^3 \,{\rho_0^3 \theta^2}/{\rho^2}\right)^{-1} \alpha_1$ } & \multicolumn{1}{c|}{0.02713926} & \multicolumn{1}{c|}{1.274657}
     \\ 
       \hline
     \multicolumn{1}{|l|}{$\tilde{\alpha}_2 \coloneqq\left(\epsilon^3\tau^3 \,{\rho_0^3 \theta^2}/{\rho^2}\right)^{-1} \alpha_2$ } & \multicolumn{1}{c|}{0.01356963} & \multicolumn{1}{c|}{0.2895342}
     \\ 
       \hline
     \multicolumn{1}{|l|}{$\tilde{\Lambda}_1 \coloneqq\left(\epsilon^3\tau^3 \,{\rho_0^3 \theta^2}/{\rho^2}\right)^{-1} \Lambda_1$ } & \multicolumn{1}{c|}{0.2137217} & \multicolumn{1}{c|}{7.534324}
     \\ 
       \hline
     \multicolumn{1}{|l|}{$\tilde{\Lambda}_2 \coloneqq\left(\epsilon^3\tau^3 \,{\rho_0^3 \theta^2}/\rho^2\right)^{-1} \Lambda_2$ } & \multicolumn{1}{c|}{0.7327600 } & \multicolumn{1}{c|}{29.93582}
     \\ 
     \hline
    \end{tabu}
    }
    \caption{Non-dimensionalized non-linear fluid parameters for Maxwell's Molecules and Hard Spheres collision operators. The viscosity parameter $\tilde{\omega}$ for the Hard Spheres operator is, up to a $0.25\%$ relative error, equal to $4/\pi$. Thus the Hard Spheres viscosity, $\omega = \tilde{\omega}\epsilon\tau \rho_0 \theta$, satisfies the well-known relation $\ell = \frac{\omega}{\rho_0}\sqrt{\frac{\pi}{2\theta}}$ for the mean free path $\ell=\epsilon L$. }
    \label{tab:nonLinearFluidParameters}
\end{table}
While the form of the fluid parameters in Table \ref{tab:nonLinearFluidParameters} is consistent with the notation used in this text, it is also useful to have the fluid parameters expressed in terms of a measurable quantity. The viscosity $\omega$ is used to accomplish this by expressing the Knudsen number in terms of $\omega$: 
$ \epsilon\tau = {\omega}/{(\tilde{\omega} \rho_0\theta)} $. 
Substituting this relation into the expressions for the other fluid parameters gives
\begin{equation}\label{eq:ESEfluidParamViscoForm}
    \begin{aligned}
      &\kappa = \frac{5}{2}\parent{\frac{\tilde{\kappa}}{\tilde{\omega}} }\, \omega \qquad \quad &\mathcal{B}= \frac{5}{2}\parent{\frac{\tilde{\mathcal{B}}}{\tilde{\omega}^3}}\, \frac{\omega^3}{\rho^2\theta^2} \qquad \quad
      &\mathcal{A} = \parent{\frac{\tilde{\mathcal{A}}}{\tilde{\omega}^3}} \, \frac{\omega^3}{\rho^2 \theta^2 }  \qquad\quad
      &\mathcal{D} = \parent{\frac{\tilde{\mathcal{D}}}{\tilde{\omega}^3}}\, \frac{\omega^3}{\rho^2\theta^2}\\
      &\mathcal{P} = \parent{\frac{\tilde{\mathcal{P}}}{\tilde{\omega}^3}}\, \frac{\omega^3}{\rho^2\theta^2}  
      &\alpha_i = \parent{\frac{\tilde{\alpha}_i}{\tilde{\omega}^3}}\, \frac{\omega^3}{\rho^2\theta}\qquad\quad
      &\Lambda_i = \parent{\frac{\tilde{\Lambda}_i}{\tilde{\omega}^3}}\, \frac{\omega^3}{\rho^2\theta}
    \end{aligned}
\end{equation}
Note that the Prandtl number is $\Pr = {\tilde{\omega}}/{\tilde{\kappa}}$. In Table \ref{tab:nonLinearFluidParametersViscForm}, the viscosity based non-dimensionalized fluid parameters are listed. We observe an interesting relationship: ${\tilde{\mathcal{A}}}/{\tilde{\mathcal{B}}} \approx \Pr^3$. For Maxwell's Molecules, this relation between $\frac{\tilde{\mathcal{A}}}{\tilde{\mathcal{B}}}$ and the Prandtl number  is actually exact while for Hard Spheres there is a $0.5\%$ relative discrepancy between the two sets of numbers. This likely has to do with the fact that the fluid parameters $\mathcal{A}$ and $\mathcal{B}$ correspond to the terms $\bm{u}\, \bm{u}\cdot\nabla[2\bm{\tepsilon(\bm{u})}]$ and $\bm{u}\, \bm{u}\cdot\nabla[\nabla\theta]$ (respectively) which are directly analogous to the Navier-Stokes-Fourier level closure terms $2\bm{\tepsilon(\bm{u})}$ and $\nabla\theta$ corresponding to the fluid parameters $\omega$ and $\kappa$. It should also be noted that some of the parameter values in Table \ref{tab:nonLinearFluidParametersViscForm} for Maxwell's Molecules and Hard Spheres are surprisingly close in value. It is also interesting to note that with the exception of $\tilde{\mathcal{A}}/\tilde{\omega}$ and $\tilde{\mathcal{B}} / \tilde{\omega}$, the Maxwell's Molecules values are larger than the Hard Spheres values.

\begin{table}[ht]
    \centering
    {\tabulinesep=2mm
    \begin{tabu}{|m{6cm}|  m{7cm}| m{7cm} | }
    \hline
    \multicolumn{1}{|c|}{\textbf{Fluid Parameters}} & \multicolumn{1}{c|}{\textbf{Maxwell's Molecules}} & \multicolumn{1}{c|}{\hspace{.2cm}\textbf{ Hard Spheres}\hspace{.4cm}}
    \\
     \hline
        \multicolumn{1}{|c|}{$\Pr $} &\multicolumn{1}{c|}{0.6666667} & \multicolumn{1}{c|}{0.6606942} 
    \\
     \hline
     \multicolumn{1}{|c|}{$\tilde{\mathcal{A}}/\tilde{\omega}^3$ } & \multicolumn{1}{c|}{0.9999998} & \multicolumn{1}{c|}{1.0419366}
     \\ 
      \hline
     \multicolumn{1}{|c|}{$\tilde{\mathcal{B}}/\tilde{\omega}^3$ } & \multicolumn{1}{c|}{3.374999} & \multicolumn{1}{c|}{3.6318842}
     \\
      \hline
     \multicolumn{1}{|c|}{$\tilde{\mathcal{D}}/\tilde{\omega}^3$ } & \multicolumn{1}{c|}{1.4999998} & \multicolumn{1}{c|}{0.9500366}
     \\
      \hline
     \multicolumn{1}{|c|}{$\tilde{\mathcal{P}} /\tilde{\omega}^3$ } & \multicolumn{1}{c|}{2.499998} & \multicolumn{1}{c|}{1.8822976}
     \\ 
      \hline
     \multicolumn{1}{|c|}{$\tilde{\alpha}_1 /\tilde{\omega}^3$ } & \multicolumn{1}{c|}{0.6666665} & \multicolumn{1}{c|}{0.6222130}
     \\ 
       \hline
     \multicolumn{1}{|c|}{$\tilde{\alpha}_2 /\tilde{\omega^3}$ } & \multicolumn{1}{c|}{0.3333333} & \multicolumn{1}{c|}{0.1413337}
     \\ 
       \hline
     \multicolumn{1}{|c|}{$\tilde{\Lambda}_1 /\tilde{\omega}^3$ } & \multicolumn{1}{c|}{5.2499997} & \multicolumn{1}{c|}{3.677816}
     \\ 
       \hline
     \multicolumn{1}{|c|}{$\tilde{\Lambda}_2 /\tilde{\omega}^3$ } & \multicolumn{1}{c|}{17.999996 } & \multicolumn{1}{c|}{14.612917}
     \\ 
     \hline
    \end{tabu}
    }
    \caption{Non-dimensionalized non-linear fluid parameters for Maxwell's Molecules and Hard Spheres collision operators when expressed in terms of the dynamic viscosity.}
    \label{tab:nonLinearFluidParametersViscForm}
\end{table}

\section{Stationary Heat Transfer Test Case}\label{sec:StationaryHeat}
\begin{figure}[h]
    \centering
    \includegraphics[width=0.3\linewidth]{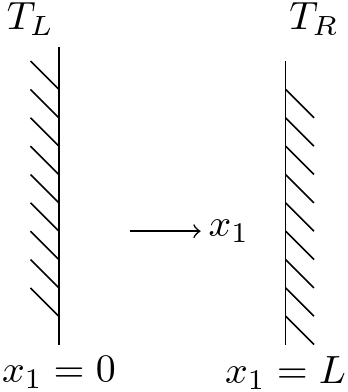}
    \caption{Setup for stationary heat transfer problem.}
    \label{fig:statHeatDomain}
\end{figure}

We first apply the quasi-linear approximation of the ESE equations\footnote{For brevity, we will refer to the quasi-linear approximation of the ESE as simply ESE for the rest of this section.} to the one-dimensional stationary heat transfer problem shown in Figure~\ref{fig:statHeatDomain}. We will assume that the walls are fully accommodating at the distribution function level. Due to this setup, we have zero bulk velocity $\bm{u} = \bm{0}$ with spatial dependence in just the $x_1$ direction. The conservation equations take the form:
\begin{subequations}\label{eq:statHeatBaseEqns}
   \begin{align}
    \dx{1}[\, \rho\theta \,] &= 0 \label{eq:statHeatBaseMomEqn}
    \\
    \dx{1}q_1 = -\dx{1}[\kappa\ \theta^2\ \dx{1}V_4] + \dx{1}^2\!\sqbrac{{(\Lambda_1+\Lambda_2)}\,\theta^2\ \dx{1}^2 V_4 } &=0 \label{eq:statHeatBaseEnergyEqn}
\end{align} 
\end{subequations}
where $V_4 = -1/\theta$. 
Notice that Equation \eqref{eq:statHeatBaseMomEqn} implies that $\rho \theta = p_0$, for some constant pressure $p_0$. We take as our reference temperature the geometric mean of the wall temperatures $\theta_0 = R\sqrt{T_L T_R}$ and define the reference density as $\rho_0 = p_0 / \theta_0$. This allows us to introduce the following non-dimensionalization:
\begin{equation*}
    \begin{aligned}
        &\hat{\rho} := \frac{\rho}{\rho_0}\qquad &\hat{\theta} := \frac{\theta}{\theta_0} = - \frac{1}{\hat{V}}  \qquad &\hat{x} := \frac{x_1}{L} \qquad &\hat{q} = \frac{q}{p_0\theta_0^{\frac{1}{2}}}
    \end{aligned}
\end{equation*}
The fluid parameters $\kappa$ and $\Lambda := \Lambda_1 + \Lambda_2$ also require non-dimensionalization. For this, we will assume a Hard Spheres collision model from now on
. Thus using Table \ref{tab:nonLinearFluidParameters}, we have 
\begin{equation*}
    \begin{aligned}
        &\kappa = \frac{5}{2} \epsilon \frac{\tilde{\omega}}{\Pr} L\ {\sqrt{\frac{\pi}{8}} }\ \rho_0 \sqrt{\theta} =  (L \rho_0\sqrt{\theta_0})\ \times  \frac{5}{2\Pr}\epsilon\,c\,\tilde{\omega} \,\hat{\theta}^{\frac{1}{2}}\\ 
        &\Lambda = \epsilon^3 (\tilde{\Lambda}_1 + \tilde{\Lambda}_2) c^3 \frac{\rho_0^3}{\rho^2} \sqrt{\theta} = (L^3 \rho_0 \sqrt{\theta_0})\ \times \epsilon^3 c^3 (\tilde{\Lambda}_1 + \tilde{\Lambda}_2)\,\hat{\theta}^{\frac{5}{2}} 
    \end{aligned}
\end{equation*}
where $c = \sqrt{\frac{\pi}{8}}$. The second expression for $\Lambda$ is obtained using $\rho\theta = \rho_0\theta_0$ to eliminate the $\rho^2$ in the denominator. Equation \eqref{eq:statHeatBaseEnergyEqn} thus takes the form
\begin{equation}\label{eq:statHeatnonDim}
    -\frac{5}{2}\frac{\epsilon\, c\, \tilde{\omega}}{\Pr}\ \frac{d}{d\hat{x}}\!\sqbrac{\hat{\theta}^{\frac{5}{2}} \frac{d\hat{V}}{d\hat{x}}} + \epsilon^2c^3 \tilde{\Lambda} \,\frac{d^2}{d\hat{x}^2}\!\sqbrac{\hat{\theta}^{\frac{9}{2}} \frac{d^2\hat{V}}{d\hat{x}^2} \,} = \ 0
\end{equation}
To fully solve this equation, we shall need two sets of boundary conditions. The form of the natural boundary conditions can be deduced from a weak form derived from \eqref{eq:statHeatnonDim}. Applying a test function $W(\hat{x})$ to \eqref{eq:statHeatnonDim} and integrating by parts over the domain $\Omega := [0, 1]$ as appropriate gives:
\begin{equation}\label{eq:statHeatESEweakFormBase}
    \begin{aligned}
        0 &= \frac{5\epsilon\,c\,\tilde{\omega}}{2\Pr}\!\int_\Omega\frac{dW}{d\hat{x}} \,
        {\hat{\theta}}^{\frac{5}{2}}\, \frac{d\hat{V}}{d\hat{x}}\,d\hat{x} + \epsilon^3 c^3\tilde{\Lambda} \int_\Omega \frac{d^2W}{d\hat{x}^2} \, {\hat{\theta}}^{\frac{9}{2}}\, \frac{d^2
        \hat{V}}{d\hat{x}^2}\, d\hat{x} 
        \\&\hspace{0.5cm}
        + \int_{\partial\Omega} W \ \hat{n}_1\! \parent{-\frac{5\epsilon\,c\,\tilde{\omega}}{2\Pr}\hat{\theta}^{\frac{5}{2}} \frac{d\hat{V}}{d\hat{x}} + \epsilon^3c^3\tilde{\Lambda} \ \frac{d}{d\hat{x}}\!\sqbrac{\hat{\theta}^{\frac{9}{2}} \frac{d^2\hat{V}}{d\hat{x}^2} } } \,dS 
        + \int_{\partial\Omega} \frac{dW}{d\hat{x}}\hat{n}_1 \parent{ - \epsilon^3c^3\tilde{\Lambda}\  {\hat{\theta}^{\frac{9}{2}}\ \frac{d^2\hat{V}}{d\hat{x}^2}}  }\, dS
    \end{aligned}
\end{equation}
where $\hat{n}_1$ is the outward pointing normal. Equation~\eqref{eq:statHeatESEweakFormBase} suggests natural boundary conditions of the form:
\begin{subequations}\label{eq:ESEbcBase}
    \begin{align}
     \hat{n}_1\parent{-\frac{5\epsilon\,c\,\tilde{\omega}}{2\Pr}\hat{\theta}^{\frac{5}{2}} \frac{d\hat{V}}{d\hat{x}} + \epsilon^3c^3\tilde{\Lambda} \ \frac{d}{d\hat{x}}\!\sqbrac{\hat{\theta}^{\frac{9}{2}} \frac{d^2\hat{V}}{d\hat{x}^2} } } &= \mathtt{h} \label{eq:ESEbcJumpBase}
     \\
     - \epsilon^3c^3\tilde{\Lambda}\  {\hat{\theta}^{\frac{9}{2}}\ \frac{d^2\hat{V}}{d\hat{x}^2}}   &= \mathtt{H}
    \label{eq:ESEbcXtraBase}
    \end{align}
\end{subequations}
for some specified functions $\mathtt{h}$ and $\mathtt{H}$.  

From the computed $\hat{V}$, we obtain the temperature $\hat{\theta}$ and the non-dimensional heat flux
\begin{equation}\label{eq:heatFluxNonDim}
    \hat{q} = -\epsilon c\,\frac{5\,\tilde{\omega}}{2\Pr}\,  \hat{\theta}^{\frac{5}{2}}
    \frac{d\hat{V}}{d\hat{x}}+ \epsilon^3 c^3 \tilde{\Lambda}\frac{d}{d\hat{x}}\!\sqbrac{\hat{\theta}^{\frac{9}{2}} \frac{d^2\hat{V}}{d\hat{x}^2} }
\end{equation} 
which takes on a constant value in the domain according to \eqref{eq:statHeatnonDim}. For any solutions generated, the heat flux at the center of the domain $(\hat{x}=0.5)$ is chosen as the representative heat flux.  

We solve for $\hat{V}$ using the following weak form: Find $\hat{V} \in\mathcal{V}$ such that for all $W\in \mathcal{V}$

\begin{equation}\label{eq:statHeatESEweakFormSolve}
    \begin{aligned}
        0 &= \frac{5\epsilon\,c\,\tilde{\omega}}{2\Pr}\!\int_\Omega\frac{dW}{d\hat{x}} \,
        {\hat{\theta}}^{\frac{5}{2}}\, \frac{d\hat{V}}{d\hat{x}}\,d\hat{x} + \epsilon^3 c^3\tilde{\Lambda} \int_\Omega \frac{d^2W}{d\hat{x}^2} \, {\hat{\theta}}^{\frac{9}{2}}\, \frac{d^2
        \hat{V}}{d\hat{x}^2}\, d\hat{x} 
        + \int_{\partial\Omega} W \ \mathtt{h} \,dS 
        + \int_{\partial\Omega} \frac{dW}{d\hat{x}}\hat{n}_1\  \mathtt{H} \ dS
    \end{aligned}
\end{equation}

For the results that follow, the test space $\mathcal{V}$ is the space of fifth degree $C^4$ B-splines on a uniform mesh of 30 elements. The weak form \eqref{eq:statHeatESEweakFormSolve} is solved using Newton's method with line search. The initial guess is
the solution to the linearized Navier-Stokes-Fourier equations with linearized Smoluchowski temperature jump boundary conditions.  All solutions were computed using the Nutils Python programming library \cite{nutils7}.

Note that to obtain the Navier-Stokes-Fourier results, the contributions due to the ESE closure are removed from Equations \eqref{eq:heatFluxNonDim} and \eqref{eq:statHeatESEweakFormSolve} or, equivalently, $\tilde{\Lambda} = \mathtt{H} = 0$. We use the (non-dimensionalized) Smoluchowski jump boundary conditions \cite{SmoluchowskivonSmolan1898,Le2012} to compute the NSF temperature and heat flux solutions used for our comparisons
\begin{equation}\label{eq:SmoluchowskiBCs}
  \parent{-\frac{5\epsilon\, c\,\tilde{\omega}}{2\Pr}\hat{\theta}^{\frac{5}{2}} \frac{d\hat{V}}{d\hat{x}}}\hat{n}_1  = \frac{4}{\sqrt{2\pi}}\,\hat{\theta}^{-\frac{1}{2}}\parent{\hat{\theta} - \frac{R T_B}{\theta_0}}
\end{equation}

Two configurations for the Stationary Heat Transfer problem are studied. The first is a moderately sized temperature difference ($T_L = 278K,\,T_R = 318K$) between the plates and the second is a large temperature difference ($T_L = 100K,\,T_R = 400K$) between the plates. For the temperature profiles, Discrete Simulation Monte Carlo (DSMC) calculations for Argon gas are used as the reference solution against which the ESE and NSF solutions generated are compared. These DSMC solutions are computed using the SPARTA DSMC code \cite{Plimpton2019}  on 120 mpi tasks in Texas Advanced Computing Center's Lonestar6 HPC system. Details on the DSMC simulation are given in Appendix~\ref{app:DSMC}. 

For heat flux comparisons, we use the Sherman-Lees \cite{Lees1965,1963rgd2.conf..228S} heat flux as the reference against which the ESE and NSF solutions are compared. To obtain this heat flux, we compute the continuum limit (i.e. $\hat{\theta}(0) = RT_L,\,\hat{\theta}(1) = RT_R$) NSF heat flux, denoted $\hat{q}_0$, and, the collisionless limit heat flux
\begin{equation}\label{eq:CollisionlessHeatFlux}
    \hat{q}^\infty = \frac{ 4 }{\sqrt{2\pi}}\frac{1}{\parent{T_LT_R}^{\frac{1}{4}}} \ \frac{T_L - T_R}{\sqrt{T_L} + \sqrt{T_R}}
\end{equation}
and use the formula
\begin{equation}\label{eq:Sherman-LeesHeatFlux}
    \frac{1}{\hat{q}^{\text{SL}}} = \frac{1}{\hat{q}^0} + \frac{1}{\hat{q}^\infty}
\end{equation}

\begin{remark}
    While each ESE and NSF solution took seconds to compute on a 2019 MacBook Pro using a 2.4GHz 8-Core Intel Core i9 processor, the DSMC solutions took between fifteen and seventy minutes to generate. This highlights the speed advantage working with moment equations have over methods such as DSMC.
\end{remark}

\subsection{The specification of $\mathtt{h}$ and $\mathtt{H}$}
In the absence of a general theory for specifying the natural boundary conditions $\mathtt{h}$ and $\mathtt{H}$, we will work with motivated ansatz. Consider the following specification:
\begin{equation}\label{eq:StatHeatBCAnsatz}
    \begin{aligned}
        \mathtt{h} &= \frac{4}{\sqrt{2\pi}}\,\hat{\theta}^{-\frac{1}{2}}\parent{\hat{\theta} - \frac{R T_B}{\theta_0}} + b_0(\epsilon)\ \hat{n}_1\hat{q}^\infty 
        \\
        \mathtt{H} &= b_1(\epsilon)\ \hat{n}_1\hat{q}^\infty 
    \end{aligned}
\end{equation}
The rationale for the above ansatz is to have the Smoluchowski jump condition dominate the specification of $\mathtt{h}$ at smaller Knudsen numbers and have the collisionless limit heat flux \eqref{eq:CollisionlessHeatFlux} influence the solution at larger Knudsen numbers through the mediation of the slip functions $b_0(\epsilon)$ and $b_1(\epsilon)$. In order to deduce the form of the slip functions, for a fixed $\epsilon_*$, we find the values that minimize the error between the ESE heat flux and the Sherman-Lees heat flux relative to the error between the NSF heat flux and the Sherman-Lees heat flux, i.e.
\begin{equation} \label{eq:optProblemStatHeat}
   \begin{pmatrix}
      b_0(\epsilon_*)\\
      b_1(\epsilon_*)
   \end{pmatrix} = \argmin_{\bm{b}\in \mathbb{R}^2} \frac{1}{2}\sum_{C_i}\frac{\parent{\hat{q}(\epsilon_*; \bm{b}, C_i) - \hat{q}^{\text{SL}}(\epsilon_*; C_i) }^{2}}{\parent{\hat{q}^{\text{NSF}}(\epsilon_*; C_i) - \hat{q}^{\text{SL}}(\epsilon_*; C_i)}^{2}} 
\end{equation}
where $C_i$ represents the temperature configurations ($T_L = 278K,\,T_R = 318K$) and ($T_L = 100K,\,T_R = 400K$), $\hat{q}^{\text{NSF}}$ is the heat flux due to NSF with the Smoluchowski boundary conditions \eqref{eq:SmoluchowskiBCs} and $\hat{q}^{\text{SL}}$ is the Sherman-Lees heat flux and $\hat{q}$ is the ESE heat flux. The details for how this inverse problem is solved are given in Appendix \ref{app:InverseProblemDetails}. 

\begin{figure}
    \centering
    \includegraphics[width=0.3\linewidth]{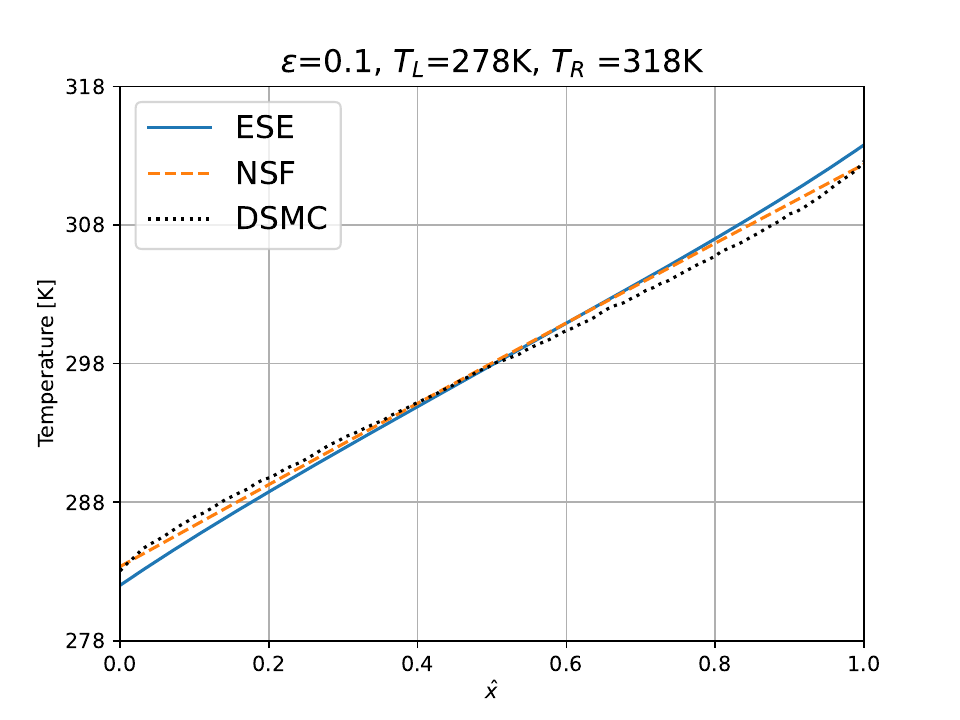}
    \includegraphics[width=0.3\linewidth]{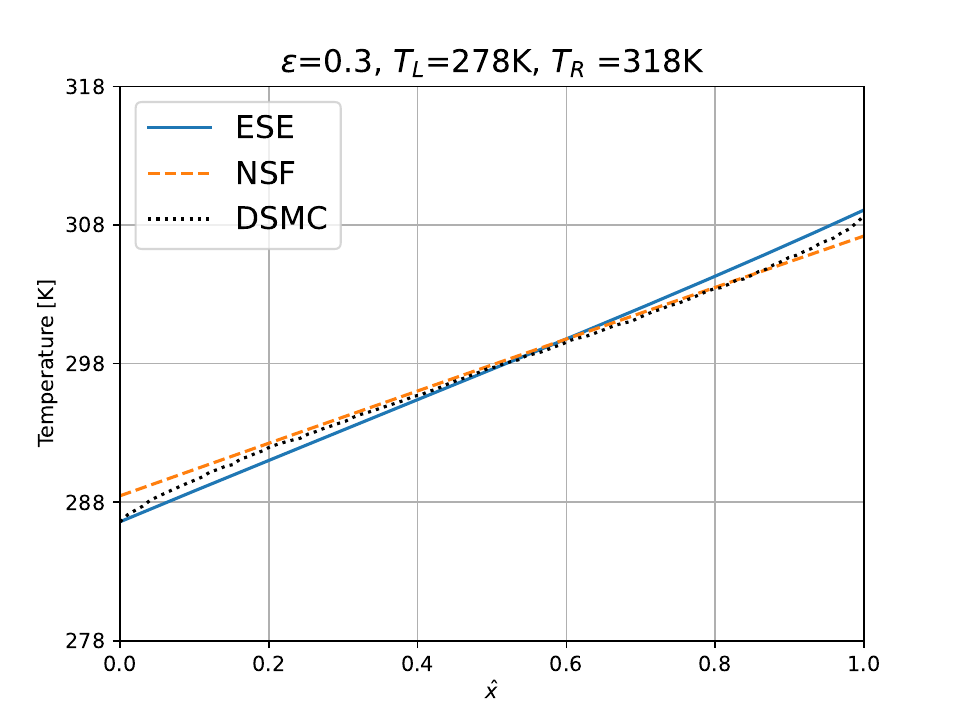}
    \includegraphics[width=0.3\linewidth]{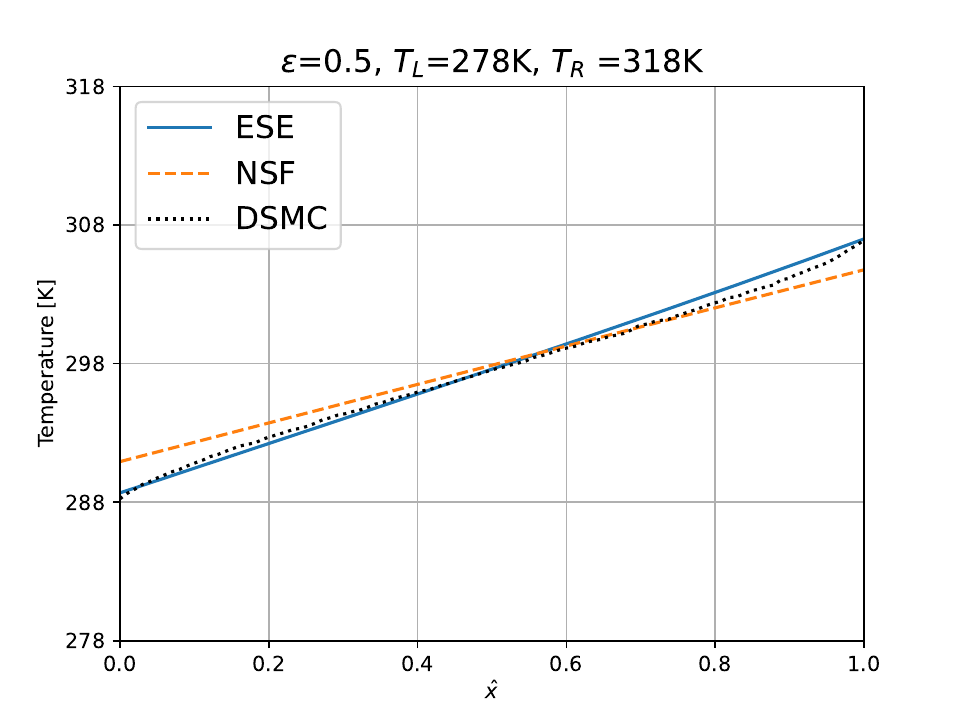}
    \includegraphics[width=0.3\linewidth]{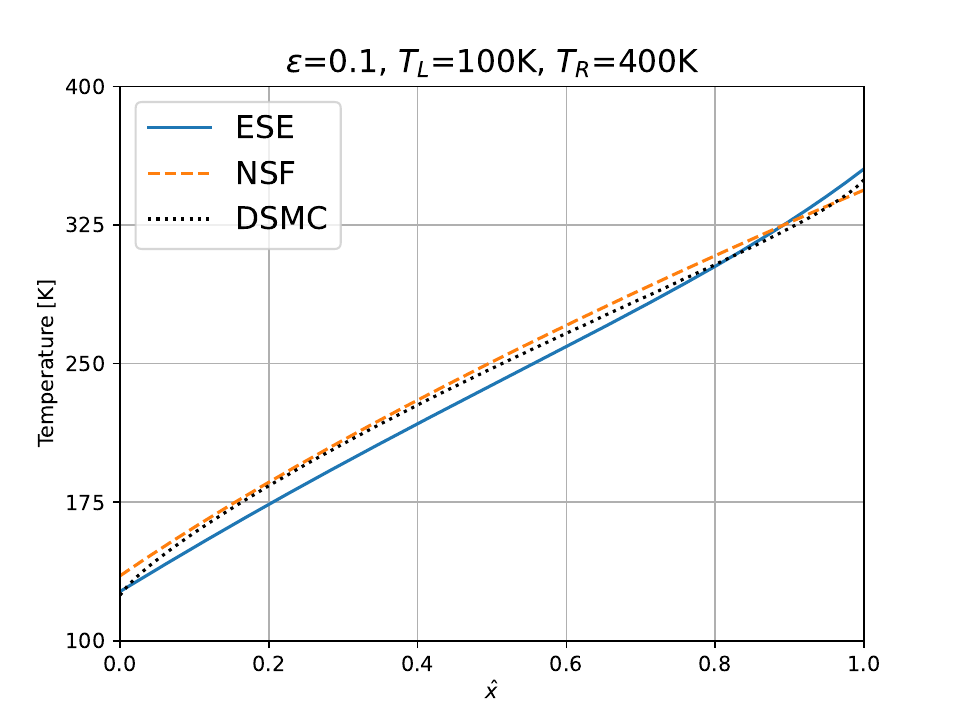}
    \includegraphics[width=0.3\linewidth]{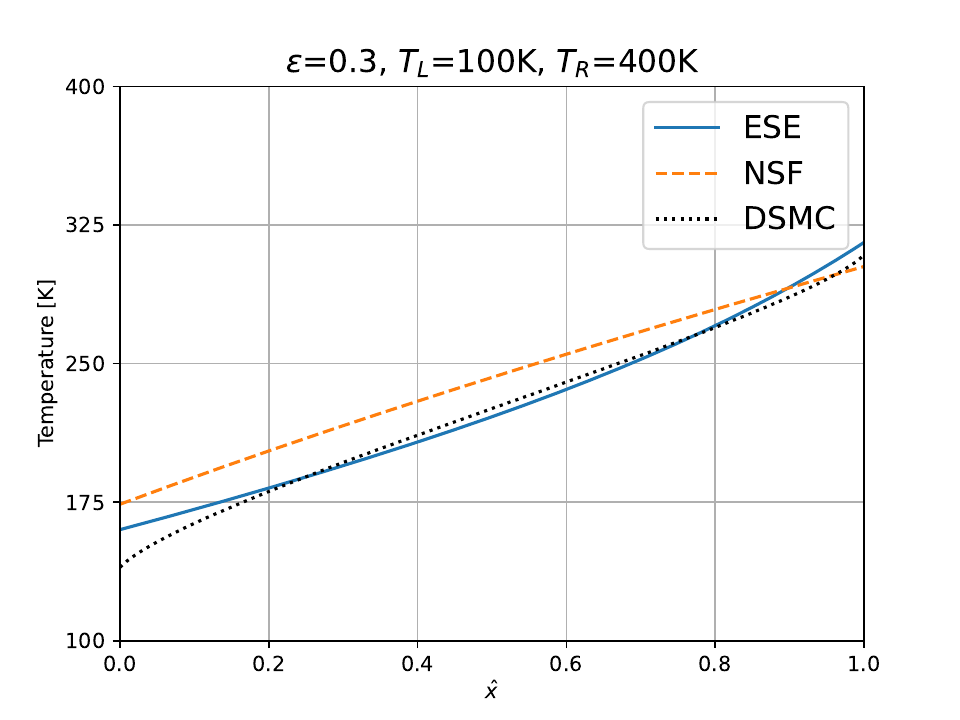}
    \includegraphics[width=0.3\linewidth]{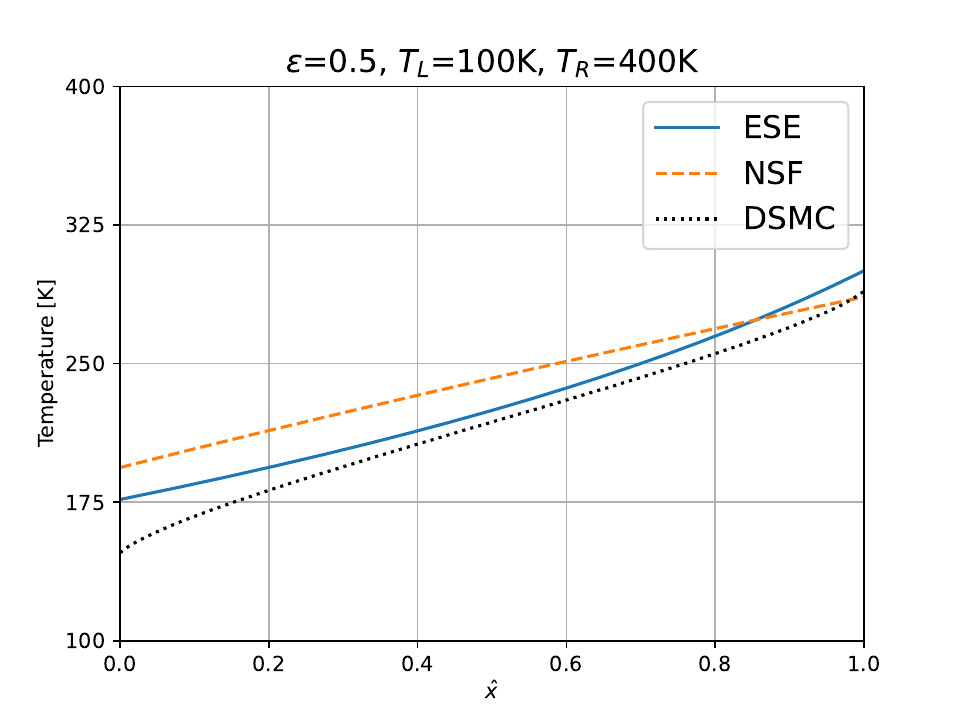}
    \caption{Temperature distribution for NSF with Smoluchowski boundary conditions \eqref{eq:SmoluchowskiBCs} and ESE with the boundary conditions \eqref{eq:StatHeatBCAnsatz} compared to a DSMC result for Argon. Knudsen numbers $\{0.1, 0.3, 0.5\}$. }
    \label{fig:TempSmallKn}
\end{figure}

\begin{figure}
    \centering
    \includegraphics[width=0.3\linewidth]{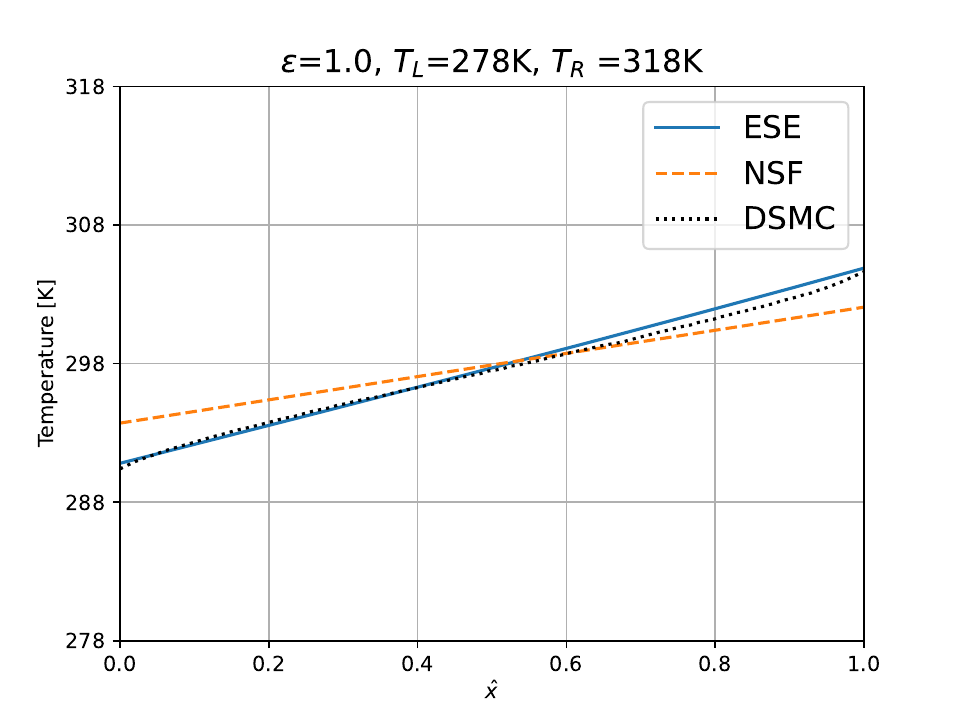}
    \includegraphics[width=0.3\linewidth]{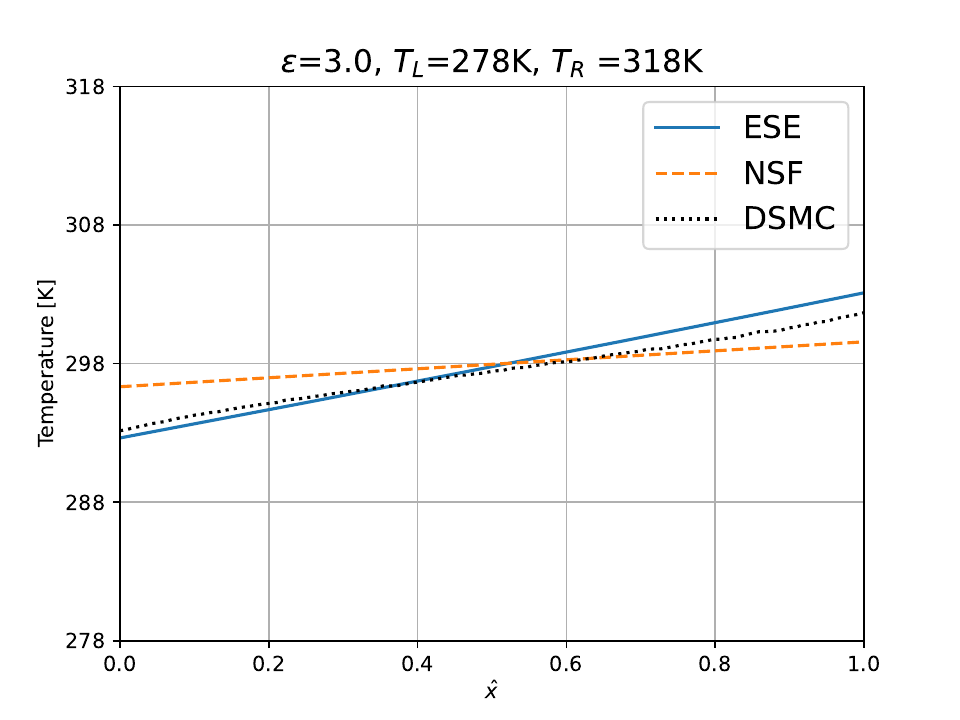}
    \includegraphics[width=0.3\linewidth]{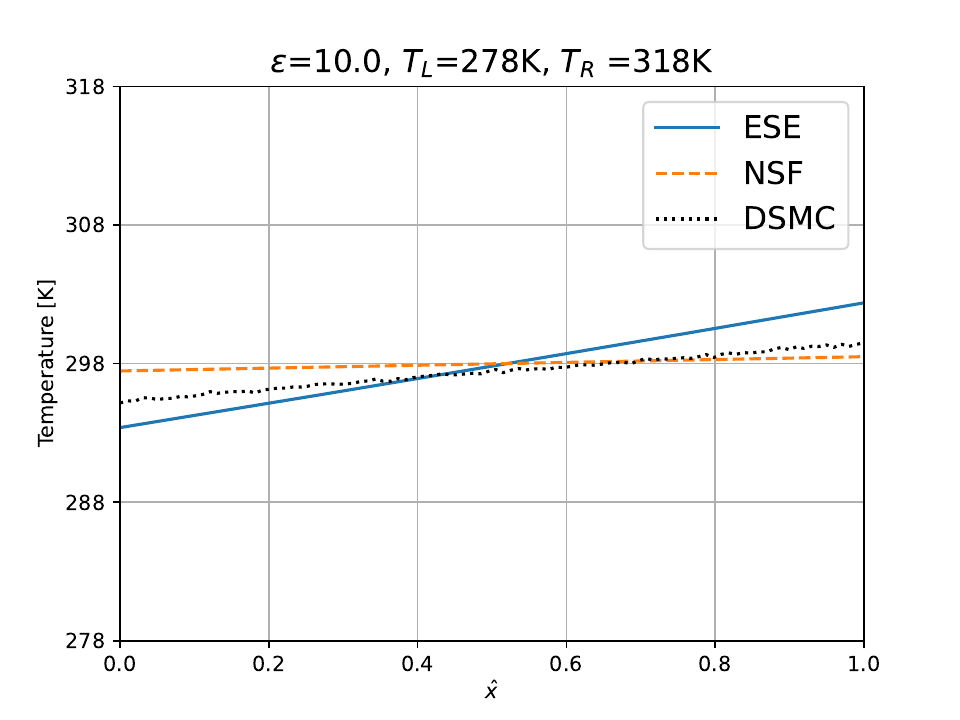}
    \includegraphics[width=0.3\linewidth]{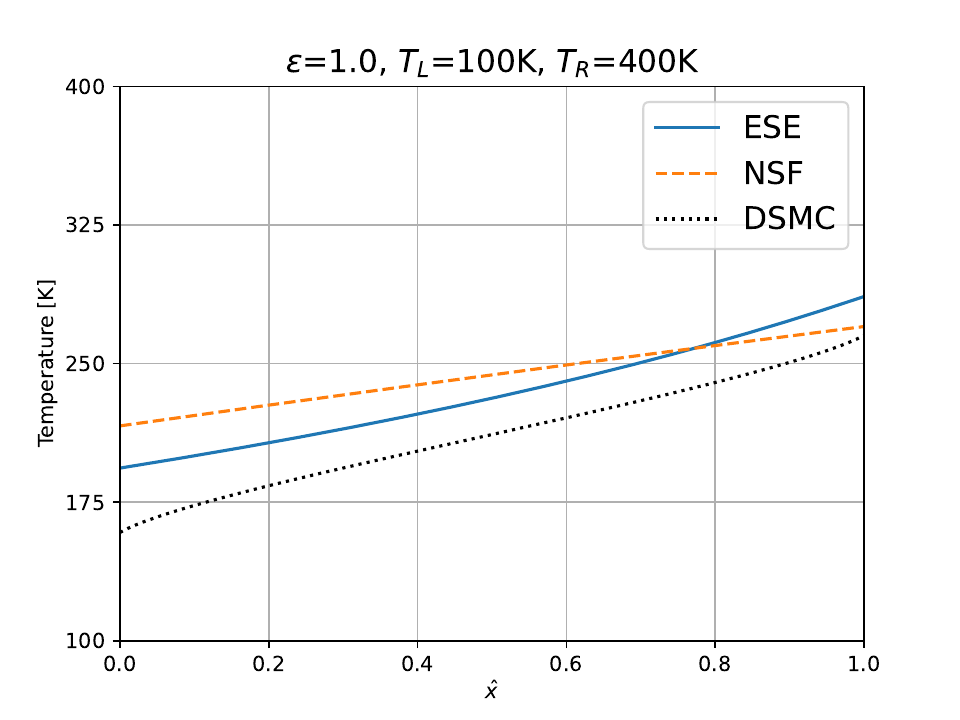}
    \includegraphics[width=0.3\linewidth]{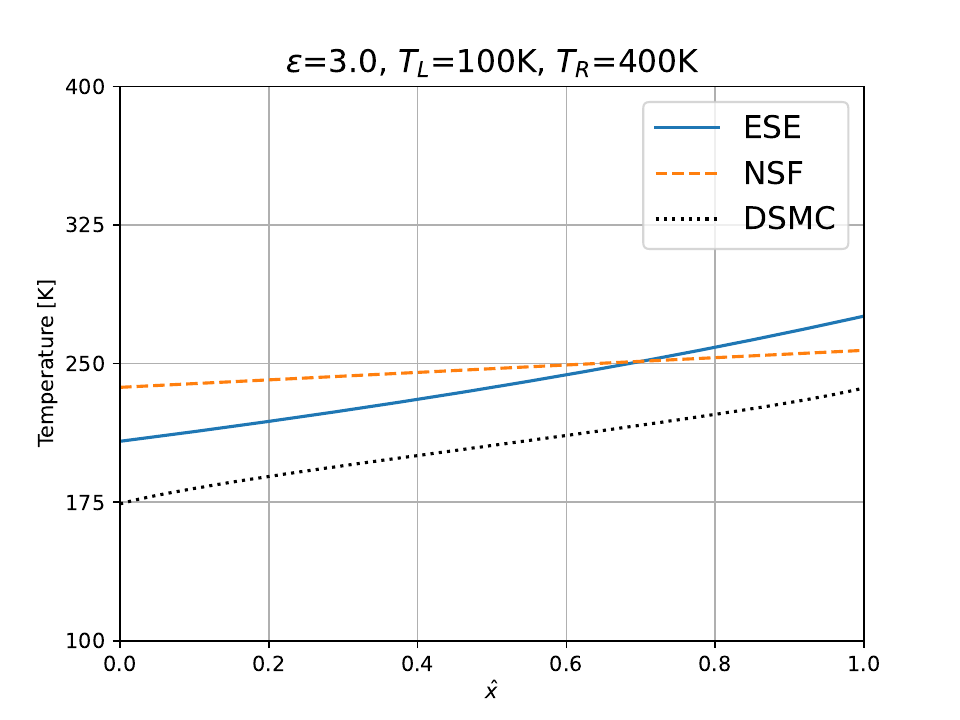}
    \includegraphics[width=0.3\linewidth]{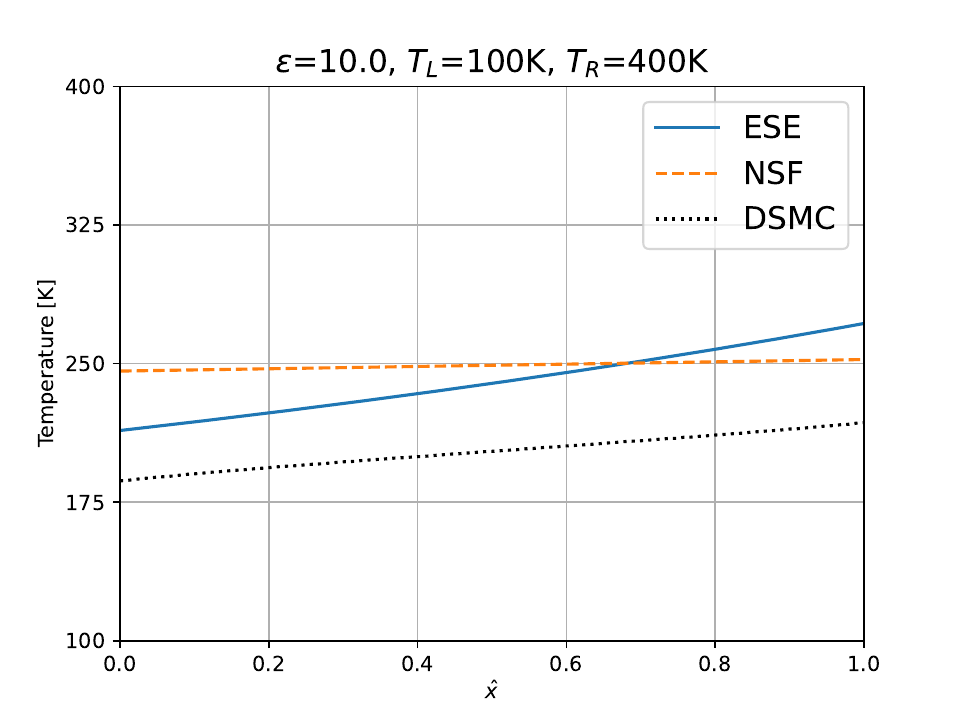}
    \caption{Temperature distribution for NSF with Smoluchowski boundary conditions \eqref{eq:SmoluchowskiBCs} and ESE with the boundary conditions \eqref{eq:StatHeatBCAnsatz} compared to a DSMC result for Argon. Knudsen numbers $\{1.0, 3.0, 10\}$ }
    \label{fig:TempBigKn}
\end{figure}

\begin{figure}
    \centering
    \includegraphics[width=0.4\linewidth]{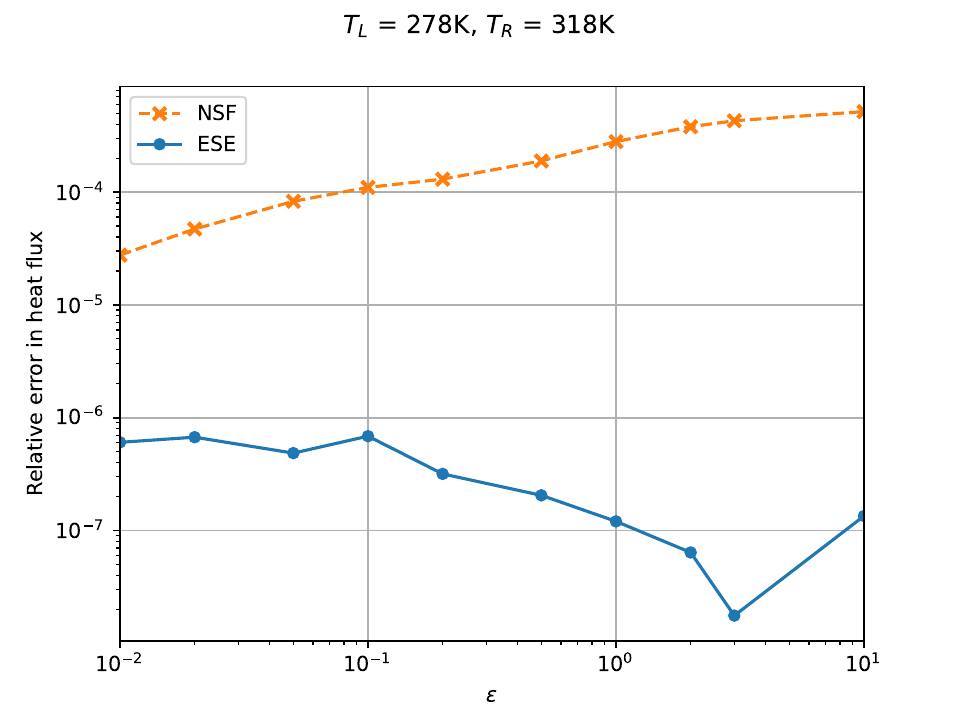}
     \includegraphics[width=0.4\linewidth]{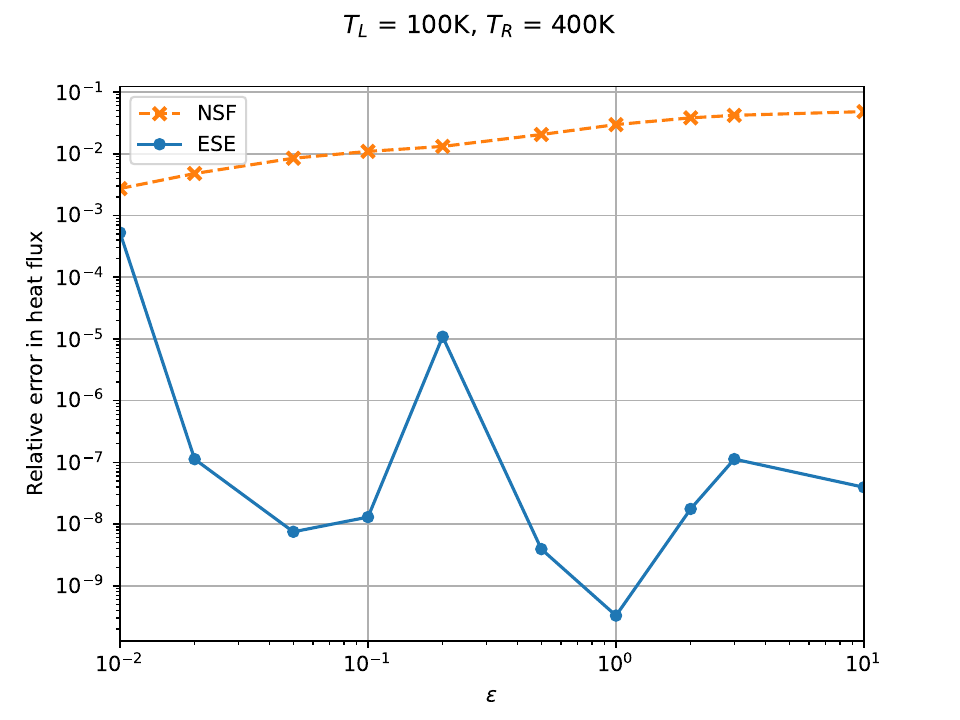}
    \caption{Relative errors $|\hat{q} - \hat{q}^{\text{SL}}|/|\hat{q}^{\text{SL}}|$ due to ESE and NSF. The choice of slip functions leads to the ESE relative heat flux error being orders of magnitude smaller than that of NSF. }
    \label{fig:heatFluxError}
\end{figure}

Figure~\ref{fig:slipFunctions} shows values of $b_0$ and $b_1$ that were obtained in this way for a range of Knudsen numbers. Figures~\ref{fig:TempSmallKn}~and~\ref{fig:TempBigKn} compares the resulting temperature distribution to the NSF and DSMC temperature distribution for a range of Knudsen numbers. Although the slip functions were chosen solely to optimize for the heat flux, we find that they tend to also yield a temperature distribution that better matches the DSMC prediction than the NSF temperature distribution does. The exception seems to be with Knudsen numbers approximately in the range $[0.05, 0.1]$, where the slip functions optimized for the heat flux results in noticeable deviation from the DSMC result compared to the NSF result. The $\epsilon=0.1$ case in Figure \ref{fig:TempSmallKn} shows what this looks like. For smaller Knudsen numbers, the ESE and NSF solutions more or less coincide. Figure \ref{fig:heatFluxError} shows that the optimized slip function values lead to a heat flux prediction at least an order of magnitude better than the NSF heat flux prediction.

\begin{figure}
    \centering
    \includegraphics[width=0.6\linewidth]{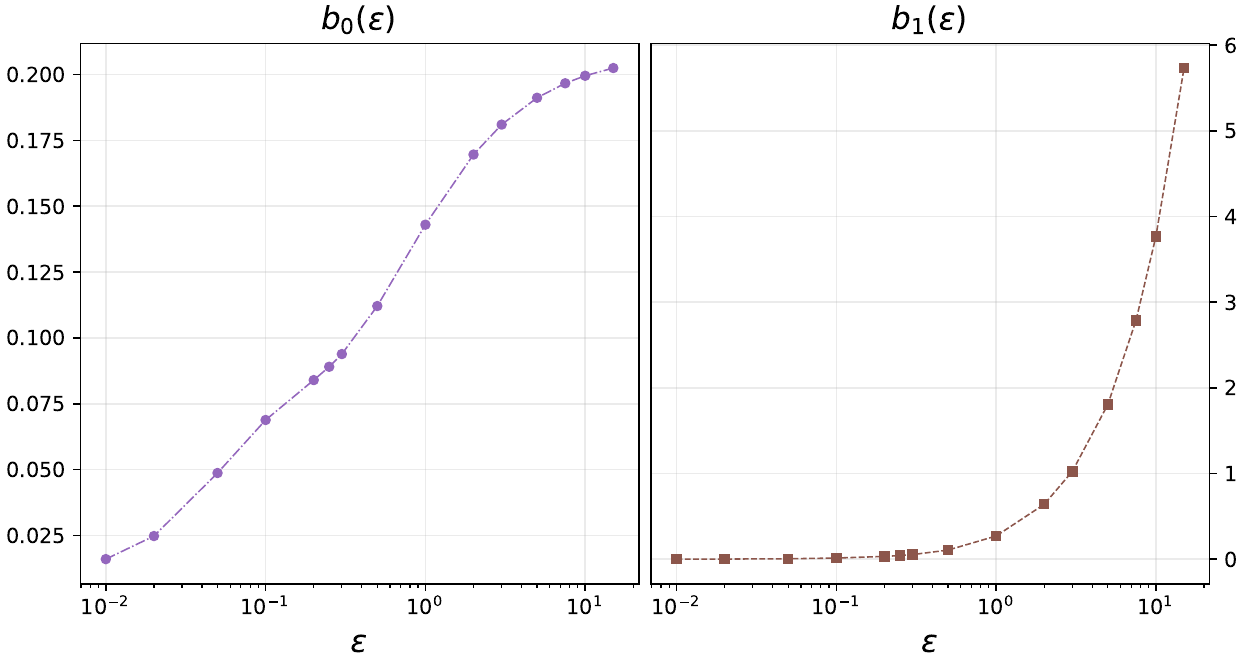}
    \caption{Values of the slip functions $b_0(\epsilon)$ and $b_1(\epsilon)$ in \eqref{eq:StatHeatBCAnsatz} that minimize \eqref{eq:optProblemStatHeat} for a given Knudsen number $\epsilon$. }
    \label{fig:slipFunctions}
\end{figure}

\section{Dispersion relations}\label{sec:Dispersion}

Consider the one-dimensional form of the equations linearized about the constant state $(\rho_0, \bm{0}, \theta_0)$:
\begin{equation}\label{eq:linESEequations}
    \begin{aligned}
        \St\partial_{t}\Tilde{\rho}+ \rho_0 \dx{1}\tu_1 &= 0 
        \\
\text{St }\rho_0\,\partial_t \tilde{u}_1 + \partial_{x_1} \!\left[\tilde{\rho}\, \theta_0 + \rho_0\, \tilde{\theta}\,\right]- \frac{4}{3}\omega\, \dx{1}^2\tu_1 + \frac{4}{9}\alpha_0\, \dx{1}^4 \tu_1 &=0 
\\
\frac{3}{2}\text{St }\,\partial_t[\rho_0 \tilde{\theta}+\tilde{\rho}\theta_0] + \frac{5}{2}\rho_0\theta_0\, \dx{1}\tu_1 - \kappa\,\dx{1}^2\ttheta + \Lambda_0 \,\dx{1}^2\ttheta &=0
    \end{aligned}
    \end{equation}
where $(\trho,\, \bm{\tu},\, \ttheta) := (\rho - \rho_0,\, \bm{u},\, \theta - \theta_0)$ are the perturbations from the constant background state. The linearized versions of the fluid parameters involve replacing $\rho$ and $\theta$ with $\rho_0$ and $\theta_0$ in the definitions given in \eqref{eq:ESEfluidParamViscoForm} and Table \ref{tab:nonLinearFluidParameters}. Furthermore, $\alpha_0 := 7 \alpha_1 + 4\alpha_2$ and $\Lambda_0 := \Lambda_1 + \Lambda_2$. 

The system of equations is non-dimensionalized using:
\begin{equation*}
    \begin{aligned}
       &\hat{\rho} = \frac{\trho}{\rho_0} \quad &\hat{u} = \frac{\tu_1}{\sqrt{\theta_0}} \quad & \hat{\theta} = \frac{\ttheta}{\theta_0}\quad &\hat{x}=
       x_1 \frac{ \rho_0}{\omega}\sqrt{\frac{2\theta_0}{\pi}}\quad &\hat{t} = \frac{t}{\St}\frac{ \rho_0\theta_0}{\omega}\sqrt{\frac{2}{\pi}}\\
    \end{aligned}
\end{equation*}
in conjunction with \eqref{eq:ESEfluidParamViscoForm}. 
The following system of equations results:
\begin{equation}\label{eq:linStabSystem}
    \begin{aligned}
     \frac{\partial \mathsf{U}}{\partial \hat{t}} + \bm{\mathsf{A}}\frac{\partial U}{\partial \hat{x}} - \bm{\mathsf{K}}\frac{\partial^2 \mathsf{U}}{\partial \hat{x}^2} + \bm{\mathsf{L}}\frac{\partial^4 \mathsf{U}}{\partial \hat{x}^4} = \mathsf{0}
    \end{aligned}
\end{equation}
where $\mathsf{U} = \big(\hat{\rho}, \hat{u}, \hat{\theta}\,\big)^{\mathsf{T}}$ and the above matrices are 
\begin{equation}\label{eq:linStabMatrices}
    \begin{aligned}
       &\bm{\mathsf{A}} = 
       \begin{pmatrix}
           0 & 1 & 0 \\
           1 & 0 & 1 \\
           0 & \frac{2}{3} & 0
       \end{pmatrix} 
       \qquad
       & \bm{\mathsf{K}} = \parent{\frac{2}{\pi}}^{\!\!\frac{1}{2}}\!
       \begin{pmatrix}
           0 & 0 & 0 \\
           0 & \frac{4}{3} & 0 \\
           0 & 0 & \frac{5}{3 \Pr}
       \end{pmatrix}
       \qquad
       & \bm{\mathsf{L}} = \parent{\frac{2}{\pi}}^{\!\!\frac{3}{2}}\!
       \begin{pmatrix}
           0 & 0 & 0 \\
           0 & \frac{4}{9}\frac{\tilde{\alpha}_0}{\tilde{\omega}^3} & 0 \\
           0 & 0 & \frac{2}{3}\frac{\tilde{\Lambda}_0}{\tilde{\omega}^3}
       \end{pmatrix}
    \end{aligned}
\end{equation}
Substituting a plane wave ansatz $\mathsf{U} = \mathsf{U}_0 \exp(i \nu \hat{t} - i k \hat{x})$ into \eqref{eq:linStabSystem} gives the
matrix equation,
$\\ (i\, \nu\, \Id_{3\times3} - i k \bm{\mathsf{A}} + k^2 \bm{\mathsf{K}} + k^4 \bm{\mathsf{L}}) \mathsf{U}_0 = \mathsf{0}  $, for which non-trivial solutions require zero determinant
\begin{equation}\label{eq:dispersionDet}
    \text{det}(i\, \nu\, \Id_{3\times3} - i k \bm{\mathsf{A}} + k^2 \bm{\mathsf{K}} + k^4 \bm{\mathsf{L}}) = 0
\end{equation}
This determinant is a third degree polynomial in $\nu$ and a fourth degree polynomial in $k^2$:
\begin{equation}\label{eq:dispersionPoly}
    \begin{aligned}
      0 &= \nu^3 - i \nu^2 \parent{\frac{4k^2}{3}\parent{\frac{2}{\pi}}^{\!\!\frac{1}{2}} + \frac{5k^2}{3\Pr}\parent{\frac{2}{\pi}}^{\!\!\frac{1}{2}} + \frac{4k^4}{9}\frac{\tilde{\alpha}_0}{\tilde{\omega}^3}\parent{\frac{2}{\pi}}^{\!\!\frac{3}{2}} + \frac{2k^4}{3}\frac{\tilde{\Lambda}_0}{\tilde{\omega}^3}\parent{\frac{2}{\pi}}^{\!\!\frac{3}{2}}}
      \\ &\hspace{.5cm}
      -\nu \parent{\frac{5k^2}{3} + \frac{20k^4}{9\Pr}\parent{\frac{2}{\pi}} + \frac{4k^6}{9}\parent{\frac{2}{\pi}}^{\!\!2}\parent{ \frac{5}{3\Pr}\frac{\tilde{\alpha}_0}{\tilde{\omega}^3} + \frac{2\tilde{\Lambda}_0}{\tilde{\omega}^3}} + \frac{8k^8}{27}\frac{\tilde{\alpha}_0\tilde{\Lambda}_0}{\tilde{\omega}^6}\parent{\frac{2}{\pi}}^{\!\!3}
      }
      \\ &\hspace{.5cm}
      +i\parent{\frac{5k^4}{3\Pr}\parent{\frac{2}{\pi}}^{\!\!\frac{1}{2}} + \frac{2k^6}{3}\frac{\tilde{\Lambda}_0}{\tilde{\omega}^3}\parent{\frac{2}{\pi}}^{\!\!\frac{3}{2}}}
    \end{aligned}
\end{equation}
Note that the corresponding polynomial equation for the NSF equations can be obtained from the above with $\tilde{\Lambda}_0 = \tilde{\alpha}_0 = 0$.



For a given real-valued wavenumber $k$, we obtain three complex-valued roots $\nu(k)$ from \eqref{eq:dispersionPoly}. After splitting into real and imaginary parts, $\nu(k) = \nu_R(k) + i\, \nu_I(k)$, observe that the imaginary part serves as the decay rate to a real-valued exponential in time, i.e. $\exp(-\nu_I(k) t)$. As such, this imaginary part of the root $\nu_I$ has to be positive for all $k$ in order for the underlying equation to be stable. The Bobylev instability observed in the Burnett equations \cite{bobylev1982chapman, Uribe2000} corresponds to them yielding a root with negative imaginary part for large enough $k$.
\begin{figure}
    \centering
    \includegraphics[width=0.495\linewidth]{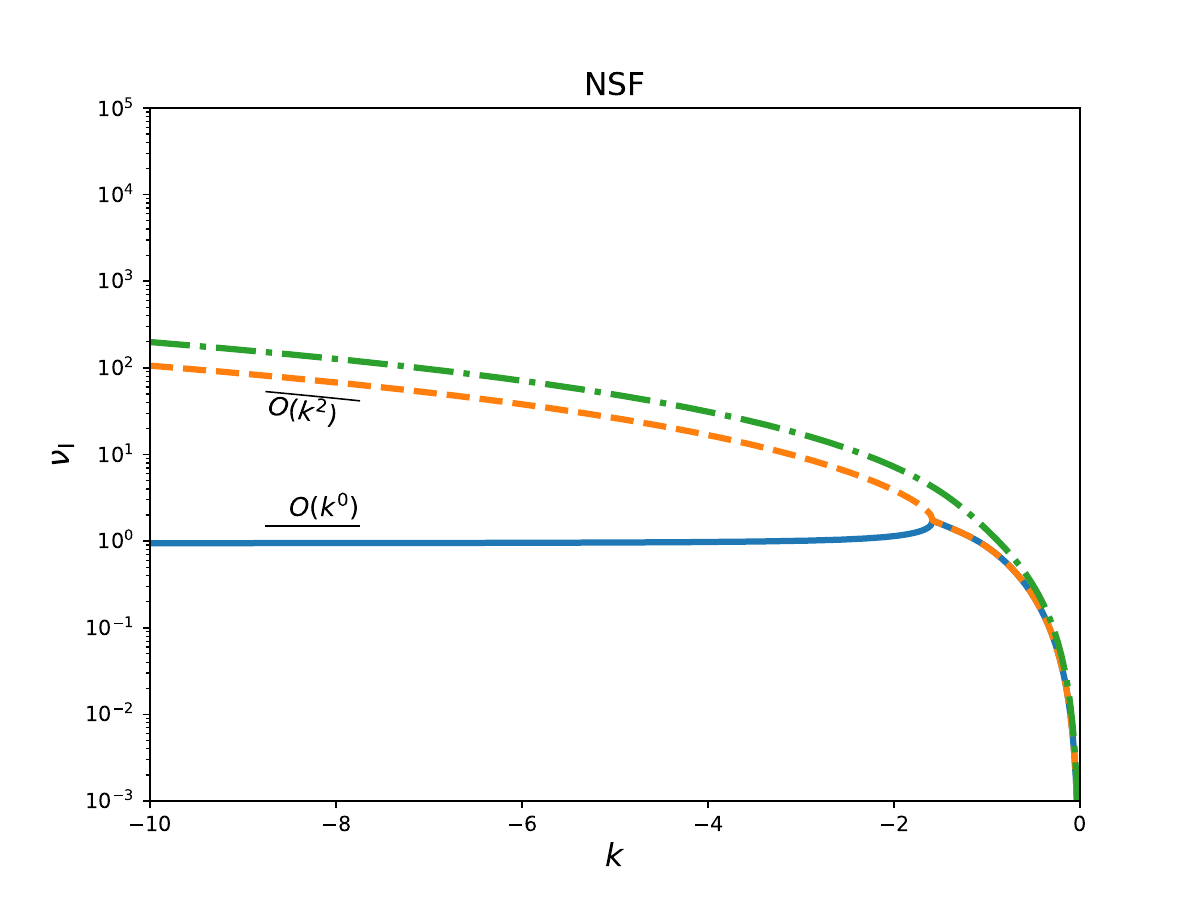}
    \includegraphics[width=0.495\linewidth]{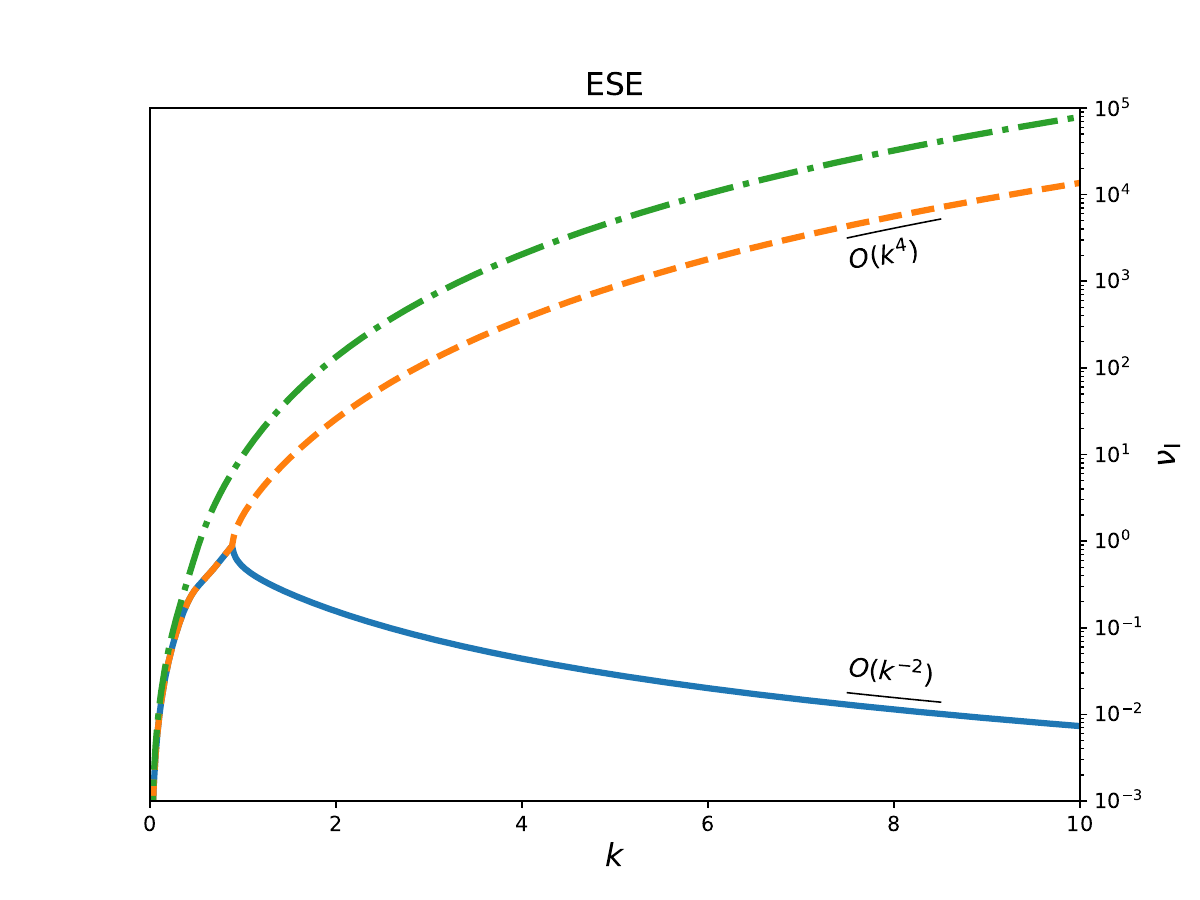}
    \caption{Plots of the imaginary part of each root, $\nu_I(k)$, as a function of $k$. The collision model used is Maxwell molecules. Note that because \eqref{eq:dispersionPoly} only contains even powers of $k$, the plots are symmetric about $k = 0$. Both NSF and ESE give stable roots. However one of the ESE roots has an imaginary part that decays as $O(k^{-2})$. }
    \label{fig:omegaDamping}
\end{figure}

 Figure \ref{fig:omegaDamping} suggests that the NSF and ESE equations are stable. That said, the ESE equations exhibits a mode that decays with growing $k^2$ (solid line in Figure \ref{fig:omegaDamping}) and it is important to show that this decaying mode does not eventually turn negative. To this end, we prove two properties of the roots of the polynomial \eqref{eq:dispersionPoly}.

 \begin{proposition}
\label{prop:realRoots} 
Suppose $\frac{2}{3}\tilde{\alpha}_0\not= \tilde{\Lambda}_0$. Then for a large enough real $k^2$, the polynomial \eqref{eq:dispersionPoly} has three distinct pure imaginary roots.
 \end{proposition}
 \begin{proof}
     It helps greatly to introduce the change of variables $\varpi = i\nu$. Equation \eqref{eq:dispersionPoly} can then be written as
\begin{equation}\label{eq:dispersionPolyRealRoots}
         0 = \varpi^3 + \varpi^2 B +\varpi C + D
     \end{equation}
      where $B,\,C,\,D$ are the corresponding terms in parenthesis in \eqref{eq:dispersionPoly} and are real-valued and positive for real-valued $k$. The proof of Proposition \ref{prop:realRoots} thus amounts to showing that \eqref{eq:dispersionPolyRealRoots} has three distinct real roots for large enough $k$. By studying the cubic polynomial discriminant  $B^2C^2 - 4C^3 -4B^3D - 27D^2 + 18BCD$, observe that for large $k^2$ the discriminant is dominated by
      $$k^{24}\! \parent{\frac{2}{\pi}}^{\!\!9} \!\parent{ \parent{\frac{4\tilde{\alpha}_0}{9\tilde{\omega}^3} + \frac{2\tilde{\Lambda}_0}{3\tilde{\omega}^3}}^{\!\!2}\parent{\frac{8}{27}\frac{\tilde{\alpha}_0\tilde{\Lambda}_0}{\tilde{\omega}^6}}^{\!\!2} - 4\parent{\frac{8}{27}\frac{\tilde{\alpha}_0\tilde{\Lambda}_0}{\tilde{\omega}^6}}^{\!\!3}} = k^{24}\! \parent{\frac{2}{\pi}}^{\!\!9}\!\parent{\frac{8}{27}\frac{\tilde{\alpha}_0\tilde{\Lambda}_0}{\tilde{\omega}^6}}^{\!\!2}\!\parent{\frac{4\tilde{\alpha}_0}{9\tilde{\omega}^3} - \frac{2\tilde{\Lambda}_0}{3\tilde{\omega}^3}}^{\!\!2}$$
If $\frac{2}{3}\tilde{\alpha}_0 \not= \tilde{\Lambda}_0$, then this term is positive and thus the discriminant is also positive. This then means that \eqref{eq:dispersionPolyRealRoots} has three distinct real roots. Thus for large enough $k^2$, the frequency roots are such that $\nu = i \nu_I$ and we can talk about the roots of \eqref{eq:dispersionPoly} and the imaginary part of the roots of \eqref{eq:dispersionPoly} interchangeably.
 \end{proof}

\begin{proposition}\label{prop:growthrates}
    For $k^2$ large enough for Proposition \ref{prop:realRoots} to hold, Equation \eqref{eq:dispersionPolyRealRoots} has two positive imaginary roots that grow as $O(k^4)$ and one positive imaginary root that decays as $O(k^{-2})$.
\end{proposition}
\begin{proof}
    Denote the roots of \eqref{eq:dispersionPolyRealRoots} by $\varpi^{(1)},\,\varpi^{(2)},\,\varpi^{(3)}$, and note that
    \begin{equation*}
        \begin{aligned}
    \varpi^{(1)}+\varpi^{(2)}+\varpi^{(3)}&= -B = O(k^4)\\
             \varpi^{(1)}\varpi^{(2)} +\varpi^{(1)}\varpi^{(3)}+ \varpi^{(2)}\varpi^{(3)} &= \ C = O(k^8)\\
             \varpi^{(1)} \varpi^{(2)} \varpi^{(3)} &= -D = O(k^6)
        \end{aligned}
    \end{equation*}
    The equation for $B$ says that at least one of the roots is negative with $O(k^4)$ growth rate. The equation for $D$ then says the other two roots are either both positive or both negative. This then means that the roots of \eqref{eq:dispersionPolyRealRoots} have a growth rate at most $O(k^4)$ because if the remaining roots have the same sign, it precludes the possibility that they grow faster than $O(k^4)$ but cancel out when added in the equation for $B$.  The equation for $C$ says there exists a combination of the products of two of the roots with growth rate $O(k^8)$. As we know there exists a negative root with $O(k^4)$ growth, this is only possible if there exists a second negative root with $O(k^4)$ growth. Finally the equation for $D$ says the product of the three roots is negative with $O(k^6)$ growth. The previous conclusions then ensure that this is only possible if the last root is negative with $O(k^{-2})$ decay.

    Finally, since $\nu = - i \varpi$, this means that the roots of \eqref{eq:dispersionPoly} are positive imaginary numbers for large enough $k^2$.
\end{proof}
 \begin{remark}
    With $\tilde{\alpha}_0 = \tilde{\Lambda}_0 = 0$ in \eqref{eq:dispersionPoly}, a similar argument to the above can be used to show that NSF will have two positive imaginary roots with $O(k^2)$ growth and one positive imaginary root with $O(k^0)$ growth for large enough $k^2$. These NSF growth rates are known (see for example, 6.12-6.14 of \cite{Sirovich1963}). 
 \end{remark}
 Figure \ref{fig:omegaDamping} shows that these theoretical growth rates manifest fully by $k^2 = 10^2$. We can therefore conclude that the ESE equations will be stable for all $k$.
 
 While still stable, the $O(k^{-2})$ decaying root of the ESE equations does present a potential issue because it means smaller wavelength perturbations are liable to decay more slowly. This could mean, for example, that due to the presence of these lingering high wavenumber transients a fine enough mesh would yield worse numerical results especially at larger Knudsen numbers (because $k$ is proportional to the Knudsen number for this problem \cite{Uribe2000}).

 \subsection{Relation between Free Wave Stability and Entropy Stability}
 The stability observed in the frequency modes is a consequence of the entropy stability of the ESE equations. To show this, first note that the linearized equations \eqref{eq:linESEequations} are entropy stable with respect to the linearized entropy $$H = \frac{1}{2}\parent{\frac{|\tilde{\rho}|^2}{\rho_0} + \frac{\rho_0}{\theta_0}|\tilde{u}_1|^2 +\frac{3\rho_0}{2}\frac{|\tilde{\theta}|^2}{\theta_0^2}} = \frac{\rho_0}{2}\parent{|\hat{\rho}|^2 +|\hat{u}|^2 + \frac{3}{2}|\hat{\theta}|^2} = \frac{\rho_0}{2}\,\mathsf{U}^* \underbrace{\begin{pmatrix}
   1&0&0\\0&1&0\\0&0&3/2  
 \end{pmatrix}}_{\bm{\mathsf{M}}} \mathsf{U}
 $$
 If we substitute the plane wave ansatz $\mathsf{U} = \mathsf{U}_0\exp(i \nu \hat{t} - ik\hat{x}) $ into \eqref{eq:linStabSystem} as before, then multiply the resulting equation by $\mathsf{U}^*\bm{\mathsf{M}}$, we obtain for real-valued $k$
 $$2\, i\nu\, H/\rho_0 - \, ik\, \mathsf{U}^*(\bm{\mathsf{M}}\bm{\mathsf{A}})\mathsf{U} = -  k^2 \mathsf{U}^*(\bm{\mathsf{M}}\bm{\mathsf{K}})\mathsf{U} - k^4 \mathsf{U}^*(\bm{\mathsf{M}}\bm{\mathsf{L}})\mathsf{U} \le 0 $$
The inequality follows because $\Pr,\, \tilde{\omega},\, \tilde{\alpha}_0,\, \tilde{\Lambda}_0$ are non-negative. The above expression is a manifestation of the entropy dissipation of the linearized ESE equations. Note too that $\mathsf{U}^*(\bm{\mathsf{M}}\bm{\mathsf{A}})\mathsf{U}$ is a real number because $\bm{\mathsf{M}}\bm{\mathsf{A}}$ is symmetric. Thus, as the right hand side is real-valued, the imaginary and real parts of the left-hand side must be such that
\begin{align*}
   &\text{Im}\big(2 \,i\nu\, H/\rho_0 - ik\, \mathsf{U}^*(\bm{\mathsf{M}}\bm{\mathsf{A}})\mathsf{U}\big) = 0 \\
   &\text{Re}(i\nu\, H) \le 0 
\end{align*}
From the second statement, we can conclude that  the imaginary part of the frequency $\nu_I$ has to always be non-negative.

\section{Conclusion}\label{sec:Conclusions}
Building on the work in \cite{Baidoo2026}, this manuscript studies the non-linear Entropy Stable Extension to the Navier-Stokes-Fourier equations (ESE) derived from the Boltzmann equation. Using entropy variables and an interesting identity related to taking the derivative of the linearized collision operator (Equation \eqref{eq:DerivativeSwitchIdentity}), the ESE is shown to take the form of a fourth-order quasi-cubic system of equations. By "quasi-cubic", we mean that the non-linearity in the derivatives of the entropy variables is cubic in nature (see Equation \eqref{eq:ESEentropyVarContribution}). We also find that the highest order derivative contributions to the ESE are contained in the quasi-linear part of the equation, which, on its own, also gives entropy stability. This allows us to isolate a structurally important, easily computed part of the ESE equations. Finally, we apply the quasi-linear approximation to the stationary heat transfer problem and then study the dispersion  of the one-dimensional linearized ESE equations.

The entropy variable description of the ESE equations has implications beyond these equations in particular. For first order and second order systems of equations such as the Euler and NSF equations, deep connections have been shown to exist between entropy functions, symmetrization and mathematical stability. The symmetrization of the ESE equations indicates that a similar connection exists for a class of non-linear fourth order system of equations. There is a promising direction for future work in determining the nature of this connection and the implications for well-posedness and numerical approximation.  

In applying the quasi-linear ESE equation to the stationary heat transfer problem, we learn that the boundary conditions specified are critical to obtaining good solutions. A key idea in this work was to specify boundary conditions such that they smoothly interpolate between the continuum limit boundary conditions and the collisionless limit boundary condition by means of a Knudsen number dependent slip function. Values for the slip functions, determined by means of an inverse problem optimizing for the heat flux (see Appendix \ref{app:InverseProblemDetails}), yielded predictions for temperature and heat flux significantly better than the corresponding predictions for the NSF equations. This indicates that the specification of boundary conditions is integral to assessing the effectiveness of a given closure. 

We believe the idea of interpolating between continuum and collisionless limits has general applicability to the specification of boundary conditions for other moment system approximations of the Boltzmann equation. That said, it has been applied in a heuristic manner in this manuscript. Questions still remain on how boundary conditions of this form could be derived from boundary conditions of the Boltzmann equation (or, at least, derived phenomenologically) and on the well-posedness of these boundary conditions. These questions are, of course, important to any class of boundary conditions applied to moment equations derived from Boltzmann and remains an active area of research. For example, there has been such work related to Onsager boundary conditions with the linearized R13 moment equations \cite{Rana2016, Hu2026}.  

It is also important to note that we only worked with the quasi-linear approximation to the ESE in this manuscript and there is still the need to compute and characterize the lower order derivative contributions due to the extra term $\mathcal{X}_{\mu}$. In problems where sharp gradients are expected, such as in shock problems, these lower order derivative contributions can be expected to play a non-negligible role. 

The stability observed when we studied the frequency roots of the dispersion polynomial \eqref{eq:dispersionPoly} is unsurprising given its relation to the entropy stability of the ESE equations. That said, the presence of the $O(k^{-2})$ decaying mode may have implications related to mesh refinement and stable discretizations of the ESE equations even though the equations are themselves stable. This merits further investigation.


\bibliography{boltzmannbook}

\appendix
\section{Appendix: The BGK collision operator}\label{app:BGK}
The Bhatnagar-Gross-Krook collision operator models collision through a relaxation to the self-consistent Maxwellian $\mu(F)$. It takes the form
\begin{equation}\label{eq:BGK}
     \mathcal{C}^{\text{BGK}}(F) = \frac{\mu(F) - F}{\tau}
\end{equation}
where $\tau$ is a relaxation timescale. For example in \cite{SoneBook, Aoki2017}, $\tau$ is given by $\tau(F) =  A_c \rho$, where $\rho = \abrac{1,F}$ and $A_c$ is a constant. The linearized BGK collision operator takes the form 
\begin{align*}
    \LM^{\text{BGK}}[g] = -\frac{1}{\tau(\mathcal{M})} (\Id - \Pi_{\mathcal{M}})[g]
\end{align*}
where $\Pi_{\mathcal{M}}$ is the orthogonal projection onto the space of collision operators. Both this collision operator and its linearized counterpart can be shown to satisfy the properties described in Sections \ref{sec:BoltzmannEquation} and \ref{sec:LinearizedCollisionOperator} with the exception of Theorem \ref{thm:LinOpDerivative} where $\mathcal{X}_{\mathcal{M}}$ no longer takes the form described by Equation \eqref{eq:XtraTermDeriv}. The fluid parameters for the BGK operator can also be computed by the procedure described in \cite{Baidoo2026, Baidoo2025}. 
The non-dimensionalized fluid parameter values for the BGK operator are:
\begin{align*}
   &\tilde{\omega} = \tilde{\kappa} = \tilde{\mathcal{A}} = \tilde{\mathcal{B}} = \tilde{\mathcal{D}} = \tilde{\mathcal{P}} = \tilde{\alpha}_1 = 1\\
   &\tilde{\alpha}_2 = 0\\
   &\tilde{\Lambda}_1 = \frac{1}{3}\\
   &\tilde{\Lambda}_2 = 9
\end{align*}

\section{Appendix: Computing the Quasi-Linear approximation and its fluid parameters}\label{app:FluidComputeProcess}
First note that 
\begin{equation}\label{eq:derivativeOflnmu}
    \dx{i}\ln\mu = \frac{\dx{i}\rho}{\rho} + (v_n - u_n)\frac{\dx{i}u_n}{\theta} + \left(\frac{|\bm{v} - \bm{u}|^2}{2} - \frac{3\theta}{2} \right) \dx{i}\!\!\sqbrac{-\frac{1}{\theta}}
\end{equation}
Thus the quasi-linear ESE term is 
\begin{equation*}
    \begin{aligned}
        \Ltens{k}{l}{j}{i}{\alpha}{\beta}(\bm{V})  \dx{j}\!\dx{i}\!V_\beta &= -\epsilon^3 \abrac{v_k\iLmu[v_l \bar{b}_\alpha], \mu \iLmu[v_j \iLmu[\vdx{i}\dx{j}\ln\mu]]}\\
        &= -\epsilon^3 \abrac{v_k\iLmu[v_l \bar{b}_\alpha], \mu \iLmu\!\!\sqbrac{v_j \iLmu\!\!\sqbrac{\dx{j}\!\!\sqbrac{\frac{\dx{i}u_n}{\theta}(v_n - u_n)v_i + \dx{i}V_4 \!\left(\frac{|\bm{v} - \bm{u}|^2}{2} - \frac{3\theta}{2} \right)v_i      }}}}
    \end{aligned}
\end{equation*}
where $V_4 = -1/\theta$ and the collision invariance property of the linearized collision operator eliminates the $\dx{i}\rho$ term. The following tensors will be useful for what comes next:
\begin{subequations}\label{eq:AlphabetTensors}
    \begin{align}
          A^\mu_{ij}(t, \bm{x}, \bm{v}) &= (\Id - \Pi_\mu)[v_i v_j] = (v_i - u_i)(v_j - u_j) - \frac{\delta_{ij}}{3}|\bm{v} - \bm{u}|^2 \label{eq:Atensor}
        \\
        B_i^\mu(t, \bm{x}, \bm{v}) &= (\Id -\Pi_\mu)\left[\frac{v_i|\bm{v}-\bm{u}|^2}{2} \right] = \frac{v_i - u_i}{2} (|\bm{v}-\bm{u}|^2 - 5\theta) \label{eq:Btensor}
        \\
        C^\mu_{ijk}(t, \bm{x}, \bm{v}) &= (v_k -u_k) \, \iLmu\big[A^\mu_{ij} \big]\label{eq:Ctensor}
        \\
        D^\mu_{ij}(t, \bm{x}, \bm{v}) &= (v_j - u_j) \, \iLmu\big[B^\mu_i \big] \label{eq:Dtensor}
    \end{align}
\end{subequations}
Taking advantage of collision invariance again gives
\begin{equation}\label{eq:ExtraTensorsBaseForm}
    \begin{aligned}
        \Ltens{k}{l}{j}{i}{\alpha}{\beta}(\bm{V})  \dx{j}\!\dx{i}\!V_\beta &= -\epsilon^3\abrac{v_k\iLmu[v_l \bar{b}_\alpha], \mu \iLmu[v_j \iLmu[A_{in}^\mu]]}\dx{j}\!\!\sqbrac{\frac{\dx{i}u_n}{\theta}}
        \\&\hspace{.5cm}
        -\epsilon^3\abrac{v_k\iLmu[v_l \bar{b}_\alpha], \mu \iLmu[v_j \iLmu[B_{i}^\mu]]}\dx{j}V_4
        \\&\hspace{.5cm}
        +\epsilon^3\abrac{v_k\iLmu[v_l \bar{b}_\alpha], \mu \iLmu[v_j \iLmu[A_{in}^\mu]]}\dx{i}V_4\, \dx{j}u_n
    \end{aligned}
\end{equation}
\begin{remark}
 We will note that in \cite{Baidoo2026, Baidoo2025}, we used $\dx{j}\dx{i}u_n/\theta$ and $\dx{j}\dx{i}\theta/\theta^2$ instead of $\dx{j}[\dx{i}u_n/\theta]$ and $\dx{j}[\dx{i}\theta/\theta^2]$ and left out the $\dx{i}V_4\, \dx{j}u_n$ term. Thus the version of the quasi-linear approximation (referred to there as the Simplified ESE) is missing several terms.   
\end{remark}  
Equation \eqref{eq:ExtraTensorsBaseForm} above gives the tensor $\mathfrak{S}_{lmk}$ with the choice $\bar{b}_\alpha = v_m$ and gives the tensor $\mathfrak{H}_{lk} + \mathfrak{S}_{lmk}u_m$ with the choice $\bar{b}_\alpha = |\bm{v}|^2/2 $. For brevity, we will focus on the derivation of $\mathfrak{S}_{lmk}$. Substituting $(\bm{v} - \bm{u}) + \bm{u} $ into \eqref{eq:ExtraTensorsBaseForm} and exploiting the rotational invariance properties of the linearized collision operator to keep only the integrands with an even number of indices leads to
\begin{equation}\label{eq:SpressCompute1}
    \begin{aligned}
        \frac{\mathfrak{S}_{lmk}}{\epsilon^3} &= -\abrac{C^\mu_{lmk}, \mu \iLmu[C^\mu_{inj}]}\parent{ \dx{j}\!\!\sqbrac{\frac{\dx{i}u_n}{\theta}} - \dx{i}V_4\dx{j}u_n}
        \\&\hspace{.5cm}
        -\abrac{A_{lm}^\mu, \Lmu^{-3}[A^\mu_{in}]}u_k\parent{\uD\!\!\sqbrac{\frac{\dx{i}u_n}{\theta}} - \dx{i}V_4\uD u_n}
        \\&\hspace{.5cm}
        -\abrac{C_{lmk}^\mu, \mu \Lmu^{-2}[B_i^\mu]}\uD \dx{i}\!V_4 - \abrac{A_{lm}^\mu, \Lmu^{-2}[D^\mu_{ij}]}u_k
        \dx{j}\dx{i}V_4
    \end{aligned}
\end{equation}
Another consequence of the rotational invariance property is that these remaining integrals are isotropic tensors. In order to fully exploit this, we introduce the following tensor 
\begin{align}\label{eq:KroneckerPhi}
        \KrPhi_{ab}^{cd} := \delta_{ac}\delta_{bd} + \delta_{ad}\delta_{bc} - \frac{2}{3}\delta_{ab}\delta_{cd}.
    \end{align}
\noindent Note the following properties: $\KrPhi_{ab}^{cd} = \KrPhi_{ba}^{cd} = \KrPhi_{ab}^{dc}$ and $\KrPhi_{kk}^{cd} = 0=\KrPhi_{ab}^{kk}$. What is more,
\begin{align*}
\KrPhi_{ab}^{cd}\ \dx{c}{w_d} &=  \dx{a}{w_b}+  \dx{b}{w_a}  - \frac{2}{3}\delta_{ab}\text{div}\bm{w}=: 2\, \tepsilon_{ab}(\bm{w}),\\
  \KrPhi_{ab}^{cd}\ \dx{c}\dx{d}g &= 2 \parent{\dx{a}\dx{b} - \frac{\delta_{ab}}{3}\Delta}\!\![\,g\,] 
\end{align*}

Because the rank 4 isotropic tensors in \eqref{eq:SpressCompute1} are symmetric and traceless in the $l,m$ indices, we can show that they evaluate to a scalar multiple of the Phi tensor $\KrPhi_{ab}^{cd}$. For example,
$$\abrac{A_{lm}^\mu, \Lmu^{-3}[D^\mu_{ij}]} = \abrac{A_{13}^\mu, \Lmu^{-3}[D^\mu_{13}]} \KrPhi_{lm}^{ij}$$
Thus we need only evaluate the integral\footnote{The integrals $\abrac{A_{12}^\mu, \Lmu^{-3}[D^\mu_{12}]}$  and $\abrac{A_{23}^\mu, \Lmu^{-3}[D^\mu_{23}]}$ work just as well.} $\abrac{A_{13}^\mu, \Lmu^{-3}[D^\mu_{13}]}$  and apply the Phi tensor to $ \dx{j}\dx{i}V_4$ in \eqref{eq:SpressCompute1}. Defining the fluid parameters
\begin{align*}
    &\mathcal{A} = -\frac{\epsilon^3}{\theta} \abrac{A_{13}^\mu, \Lmu^{-3}[A^\mu_{13}]}
    &\mathcal{P} = -\frac{\epsilon^3}{\theta}\abrac{C_{131}^\mu, \mu \Lmu^{-2}[B_3^\mu]}\hspace{1.cm}
    &\mathcal{D} = -\frac{\epsilon^3}{\theta}\abrac{A_{13}^\mu, \Lmu^{-2}[D^\mu_{13}]}
\end{align*}
and applying the Phi tensor to the corresponding terms in \eqref{eq:SpressCompute1} leads to the $\mathcal{A}$, $\mathcal{P}$ and $\mathcal{D}$ terms in \eqref{eq:ESESpress}.

This leaves the more complicated contribution due to the $K_{lmk}^{inj}:= \abrac{C^\mu_{lmk}, \mu \iLmu[C^\mu_{inj}]}$ term. Exploiting the symmetry and traceless properties of the $l,m$ and $i,n$ indices pairs, one can show that
\begin{align*}
   K_{lmk}^{inj} &= K_{123}^{123}\parent{\delta_{kj}\KrPhi_{lm}^{in}+\delta_{ki}\KrPhi_{lm}^{nj}+ \delta_{kn}\KrPhi_{lm}^{ij}-\frac{2}{3}\delta_{in}\KrPhi_{lm}^{kj}} \\
   &\hspace{.5cm}
   +K_{122}^{133}\parent{\delta_{ij}\KrPhi_{lm}^{nk}+\delta_{jn}\KrPhi_{lm}^{ik}-\frac{2}{3}\delta_{in}\KrPhi_{lm}^{kj} }
\end{align*}
By defining the fluid parameters
\begin{align*}
    &\alpha_1 = -\frac{\epsilon^3 K_{123}^{123}}{\theta}  &\alpha_2 = -\frac{\epsilon^3 K_{122}^{133}}{\theta}
\end{align*}
and applying the above isotropic tensor decomposition to the term $(\dx{j}[\dx{i}u_n/\theta] - \dx{i}V_4 \dx{j}u_n)$ then gives the corresponding $\alpha_1$ and $\alpha_2$ contributions in \eqref{eq:ESESpress}.

The computation of the fluid parameters rely on a Burnett function decompositions of the tensors \eqref{eq:AlphabetTensors} and the use of collision matrices that result from applying the linearized collision operator to the Burnett functions. Details for that computation can be found in \cite{Baidoo2026, Baidoo2025}.

\section{Appendix: DSMC simulation of stationary heat transfer}\label{app:DSMC}

The DSMC temperature results shown in Figures \ref{fig:TempSmallKn} and \ref{fig:TempBigKn} were obtained through a DSMC simulation carried out in a rectangular box of $(5 \times 5 \times 500)\mu \text{m}$. Periodic boundary conditions were set on the boundaries perpendicular to the $x$ and $y$ axes and the boundaries perpendicular to the $z$ axis were fully accommodating with the temperatures $T_L$ and $T_R$. The box was divided into 100 uniform square cells. This ensured that each cell was less than a third the length of the mean free path. The particles used in the domain were Argon gas with the Variable Soft Spheres collision model $(d_{\text{ref}} = 3.657897\times10^{-10}\text{m},\, \text{omega} = 0.81,\, T_{\text{ref}} = 273.15 K, \, \text{alpha} = 1.0)$.

The simulation was carried out using 120 mpi tasks on Texas Advanced Computing Center Lonestar6 HPC system. The timestep used was 
$$\Delta t = \frac{\ell}{32}\sqrt{\frac{\pi}{4(T_L + T_R)R_{\text{Ar}}}}$$
where $\ell = 500\epsilon\,\mu\text{m}$ is the mean free path and $R_{\text{Ar}} = 208.13 \text{J/kg}K$ is the specific gas constant of Argon. The ratio of real to simulated particles used (\textit{fnum}), and the number of time steps before results were recorded are given below.
\begin{table}[ht]
    \centering
    {\tabulinesep=2mm
    \begin{tabu}{|m{7cm}|  m{7cm}|  m{7cm}|  m{7cm}|  m{7cm}|  m{7cm}|  m{7cm}| }
    \hline
    \multicolumn{1}{|c|}{$\epsilon$} &  \multicolumn{1}{c|}{0.1} & \multicolumn{1}{c|}{0.5} & \multicolumn{1}{c|}{1.0} & \multicolumn{1}{c|}{2.0}
     & \multicolumn{1}{c|}{3.0}
      & \multicolumn{1}{c|}{10.0}
      \\
    \hline
    \multicolumn{1}{|c|}{\textit{fnum}}
    & \multicolumn{1}{c|}{$10^4$}
     & \multicolumn{1}{c|}{$10^4$}
     & \multicolumn{1}{c|}{$10^3$}
     & \multicolumn{1}{c|}{$10^4$}
     & \multicolumn{1}{c|}{$10^3$}
     & \multicolumn{1}{c|}{$10^3$}\\
    \hline
    \multicolumn{1}{|c|}{total timesteps}
    & \multicolumn{1}{c|}{$5\times 10^6$}
     & \multicolumn{1}{c|}{$5 \times 10^6$}
     & \multicolumn{1}{c|}{$2 \times 10^6$}
     & \multicolumn{1}{c|}{$2 \times 10^6$}
     & \multicolumn{1}{c|}{$1 \times 10^6$}
     & \multicolumn{1}{c|}{$5 \times 10^5$}\\
     \hline
    \end{tabu}
    }
    \label{tab:SPARTAparams}
\end{table}
The seed used for the simulations was $\mathtt{12345}$. 

\section{Inverse Problem for boundary slip functions}\label{app:InverseProblemDetails}
Recall that the problem of interest is 
\begin{equation} \label{eq:APPoptProblemStatHeat}
   \begin{pmatrix}
      b_0(\epsilon_*)\\
      b_1(\epsilon_*)
   \end{pmatrix} = \argmin_{\bm{b}\in \mathbb{R}^2} \frac{1}{2}\sum_{C_i}\frac{\parent{\hat{q}(\epsilon_*; \bm{b}, C_i) - \hat{q}^{\text{SL}}(\epsilon_*; C_i) }^{2}}{\parent{\hat{q}^{\text{NSF}}(\epsilon_*; C_i) - \hat{q}^{\text{SL}}(\epsilon_*; C_i)}^{2}} =: \argmin_{\bm{b}\in\mathbb{R}^2}\, \mathcal{S}(\epsilon_*; \bm{b})
\end{equation}
The minimization problem \eqref{eq:APPoptProblemStatHeat} can be formulated in terms of finding a point $\bm{b}_*$ such that
\begin{equation*}
    \bm{0} =  \frac{\partial \mathcal{S}}{\partial\bm{b}}(\epsilon_*; \bm{b}_*)
\end{equation*}
This point is found using Newton's method
\begin{equation}\label{eq:NewtonIteration}
    \bm{b}^{n+1} = \bm{b}^n - \parent{ \frac{\partial^2\mathcal{S}}{\partial \bm{b} \partial \bm{b} }(\epsilon_*;\bm{b}^n)}^{\!\!-1} \parent{ \frac{\partial\mathcal{S}}{\partial \bm{b} }(\epsilon_*;\bm{b}^n)}
\end{equation}
where 
\begin{equation*}
    \begin{aligned}
      \frac{\partial\mathcal{S}}{\partial \bm{b} }(\epsilon_*;\bm{b}^n) &= \sum_{C_i} \frac{\hat{q} - \hat{q}^{\text{SL}}}{\parent{\hat{q}^{\text{NSF}} - \hat{q}^{\text{SL}}}^{2}} \frac{\partial \hat{q}}{\partial \bm{b}} \\
      \frac{\partial^2\mathcal{S}}{\partial \bm{b} \partial \bm{b} }(\epsilon_*;\bm{b}^n) &= \sum_{C_i} \frac{ 1}{\parent{\hat{q}^{\text{NSF}} - \hat{q}^{\text{SL}}}^{2}} \parent{ \big(\hat{q} - \hat{q}^{\text{SL}}\big) \frac{\partial^2 \hat{q}}{\partial \bm{b} \partial \bm{b}} 
      +  \, \frac{\partial \hat{q}}{\partial \bm{b}} \otimes \frac{\partial \hat{q}}{\partial \bm{b}}}
    \end{aligned}
\end{equation*}

The derivatives $\partial\hat{q}/\partial \bm{b}$ and $\partial^2\hat{q}/\partial \bm{b} \partial\bm{b}$ are obtained by solving for $\partial\hat{V}/\partial \bm{b}$ and $\partial^2\hat{V}/\partial \bm{b} \partial\bm{b}$ using the weak form \eqref{eq:statHeatESEweakFormSolve} with our specification \eqref{eq:StatHeatBCAnsatz}:
\begin{equation*}
\begin{aligned}
  0= \text{res}\big(W, \hat{V}(\bm{b}); \epsilon, \bm{b}, C_i \big) &:= \frac{5\epsilon\,c\,\tilde{\omega}}{2\Pr}\!\int_\Omega\frac{dW}{d\hat{x}} \,
        {\hat{\theta}}^{\frac{5}{2}}\, \frac{d\hat{V}}{d\hat{x}}\,d\hat{x} + \epsilon^3 c^3\tilde{\Lambda} \int_\Omega \frac{d^2W}{d\hat{x}^2} \, {\hat{\theta}}^{\frac{9}{2}}\, \frac{d^2
        \hat{V}}{d\hat{x}^2}\, d\hat{x} 
        \\&\hspace{0.5cm}
        + \int_{\partial\Omega} W \parent{\frac{4}{\sqrt{2\pi}}\,\hat{\theta}^{-\frac{1}{2}}\parent{\hat{\theta} - \frac{R\, T_B}{\theta_0}} + b_0\ \hat{n}_1\hat{q}^\infty } \,dS 
        + \int_{\partial\Omega} \frac{dW}{d\hat{x}}\,b_1\ \hat{q}^\infty \ dS    
\end{aligned}   
\end{equation*}
After solving for $\hat{V}(\bm{b^n})$ in the above, we find $\frac{\partial\hat{V}}{\partial \bm{b}}(\bm{b}^n)$ by solving the linear equations
\begin{equation*}
    0 = \frac{\partial}{\partial {b}_k}\text{res}\big(W, \hat{V}(\bm{b}^n); \epsilon, \bm{b}^n, C_i \big) = \left.\frac{d}{d\eta} \text{res}\!\left(W, \hat{V}(\bm{b}^n) + \eta \frac{\partial\hat{V}}{\partial {b}_k}; \epsilon, \bm{b}^n, C_i \right)\right|_{\eta=0} + \text{res}\left(W, \hat{V}(\bm{b}^n); \epsilon, \hat{\bm{d}}_k, C_i \right) 
\end{equation*}
where $\hat{\bm{d}}_0 = (1, 0)^T$ and $\hat{\bm{d}}_1 = (0, 1)^T$.
Finally, with $\hat{V}(\bm{b^n})$ and $\frac{\partial\hat{V}}{\partial \bm{b}}(\bm{b}^n)$ in hand, we calculate $\frac{\partial^2\hat{V}}{\partial \bm{b} \partial \bm{b}}(\bm{b}^n)$ using the linear problem
\begin{equation*}
    0 = \frac{\partial^2}{\partial b_i \partial b_j}\text{res}\big(W, \hat{V}(\bm{b}^n); \epsilon, \bm{b}^n, C_i \big) 
\end{equation*}
For these linearizations, it is important to note that $\frac{\partial}{\partial\bm{b}}\big[\hat{\theta}^r \big] = r\, \hat{\theta}^{r+1}\, {\partial \hat{V}}/{\partial \bm{b}}$.
Taking derivatives of \eqref{eq:heatFluxNonDim} then gives $\partial\hat{q}/\partial \bm{b}$ and $\partial^2\hat{q}/\partial \bm{b} \partial\bm{b}$.

\end{document}